\documentclass[prb,aps,twocolumn,floatfix]{revtex4-1} 
\usepackage{epsf}
\usepackage{epsfig}
\usepackage{graphics}
\begin{document}

\title{
A real space auxiliary field approach to the BCS-BEC crossover
}

\author{Sabyasachi Tarat and Pinaki Majumdar}

\affiliation{Harish-Chandra  Research Institute,
Chhatnag Road, Jhusi, Allahabad 211019, India}

\date{2 Feb 2014}

\begin{abstract}
The BCS to BEC crossover in attractive Fermi systems is a 
prototype of weak to strong coupling evolution in many body
physics. While extensive numerical results  are available,
and several approximate methods have been developed, most of 
these schemes  are unsuccessful in the presence of spatial
inhomogeneity. Such situations call for a real space approach
that can handle large spatial scales and retain the crucial 
thermal fluctuations. With this in mind, we present
comprehensive results of a real space auxiliary field approach 
to the BCS to BEC crossover in the attractive Hubbard model in 
two dimensions. The scheme reproduces 
the Hartree-Fock-Bogoliubov ground state, and leads to a $T_c$
scale that agrees with quantum Monte Carlo estimates to within 
a few percent.  We provide results on the $T_c$, amplitude and 
phase fluctuations, density of states, and the momentum resolved 
spectral function over the entire interaction and temperature 
window. We suggest how the method generalises successfully to 
the presence of disorder, trapping, and population imbalance.
\end{abstract}

\maketitle

\section{Introduction}

The BCS to BEC crossover in attractive fermion systems is
a topic of enduring interest \cite{bcs-bec-rev}.
With increasing interaction, the ground state of a weak coupling 
`BCS superconductor' \cite{bcs}, 
with pair size $\xi$ 
much larger than the interparticle separation $k_F^{-1}$,
evolves smoothly \cite{eagles,leggett,noz,micnas,rand-rev}
into a `Bose-Einstein condensate' (BEC)
of preformed fermion pairs with $\xi \lesssim  k_F^{-1}$.
$k_F$, above, is the Fermi wavevector.
The `high temperature' normal state changes from a 
conventional Fermi liquid at weak coupling to a gapped phase 
at strong coupling. 
While the zero temperature 
pairing gap increases with coupling strength,  
the superconducting $T_c$ in lattice systems reaches a
maximum at intermediate coupling and falls thereafter.
A striking consequence of the separation of
pairing and superconducting scales is the emergence of
a (pseudo)gapped normal phase, 
with preformed fermion pairs but no
superconductivity  due to strong phase fluctuations.

The early work of Leggett \cite{leggett}
and Nozieres  and Schmitt-Rink \cite{noz} provided the intuitive
basis for understanding this problem. It has since been 
followed up by 
powerful semi-analytic schemes 
\cite{sa-2psc,sa-deisz,sa-kopec,sa-dupuis,sa-miyake}, 
extensive quantum Monte Carlo (QMC) work 
\cite{qmc-scal1,qmc-moreo1,qmc-moreo2,qmc-spingap,qmc-nfl,qmc-singer,qmc-trm1,qmc-beck,qmc-paiv1,qmc-paiv2}, 
and most recently
dynamical mean field theory (DMFT) 
\cite{dmft-keller,dmft-castel,dmft-garg,dmft-bauer,dmft-dup,dmft-koga}.
The efforts have established the non-monotonic $T_c$,
 and the presence of a pseudogap   
in the single particle spectrum beyond moderate coupling
and temperature $T > T_c$.

While the success of multiple methods in capturing the 
crossover is remarkable, most of them depend on
translation invariance in the
underlying problem. 
They do not naturally generalise to 
problems that involve the presence of disorder,
or a confining potential, or the emergence of spontaneous
modulation.
These situations, for example, occur in the 
context of the disorder driven superconductor-insulator 
transition \cite{sit-rev},
trapped fermions in an optical lattice \cite{opt-rev},
and Fulde-Ferrell-Larkin-Ovchinnikov (FFLO) states \cite{fflo-rev}
in population imbalanced systems.
Such problems, in general, require a real space 
approach.
This paper presents such a real space
implementation based
on a static auxiliary field (SAF) scheme.

Our main results are the following:
(i)~We demonstrate the ability of our approach to 
quantitatively capture the $T_c$ scale across the BCS-BEC
crossover, confirming its usefulness at all interaction
strengths.
(ii)~We quantify the crossover from an amplitude
fluctuation dominated regime to a phase fluctuation
dominated regime, through the `high $T_c$' 
intermediate coupling window where both are important.
(iii)~We present the thermal evolution of the single
particle density of states with interaction and 
temperature, and, more importantly, the momentum
resolved spectral function $A({\bf k},\omega)$.
We compare our results to QMC data 
wherever available.

The paper is organised as follows. 
In Sec.II we quickly compare the existing analytic and
numerical methods used to study the BCS-BEC crossover in
lattice models. Sec.III presents our model and 
describes the method used in detail. Sec.IV shows our
results on thermodynamic indicators, the nature of
fluctuations, density of states, and $A({\bf k},\omega)$.
Sec.V discusses the limitations of our method and
the scope for further work.

\section{Earlier work}

Since the BCS-BEC crossover is a prototype of
weak to strong coupling evolution, several methods,
of increasing sophistication, have been brought to
bear on it.
These include mean field theory (MFT)
\cite{leggett}, MFT corrected by 
gaussian fluctuations \cite{noz},
the self consistent T-matrix approach (SC-TMA)
\cite{sa-deisz,sa-miyake},
a two particle self consistent (2PSC) scheme 
\cite{sa-2psc}, the mapping to XY models
\cite{sa-dupuis,sa-kopec},
quantum Monte Carlo (QMC)
\cite{qmc-scal1,qmc-moreo1,qmc-moreo2,qmc-spingap,qmc-nfl,qmc-singer,qmc-trm1,qmc-beck,qmc-paiv1,qmc-paiv2}, 
and recently dynamical mean field theory
(DMFT) 
\cite{dmft-keller,dmft-castel,dmft-garg,dmft-bauer,dmft-dup,dmft-koga}.

There are detailed descriptions of these methods available
in the original literature so we just provide a table that
compares the strengths and limitations of these methods 
in the light of a few crucial indicators. These in our
opinion include (i)~thermodynamics: 
$T_c(U)$, (ii)~single particle spectra, 
(iii)~two particle properties, {\it e.g},
conductivity, and 
(iv)~handling inhomogeneity, {\it e.g}, 
disorder or trapping.
%\newpage
\begin{table*}
 
\begin{center}
\begin{tabular}{ | p{3.0cm} | p{3.5cm} | p{3.5cm} | p{3.5cm} | p{3.9cm} | } 
\hline    
Method & ~~~~~~~~~ $T_{c}(U)$~~~~~~~~~~~~ & Spectra at large $U/t$  & Transport at large $U/t$ & 
Handling inhomogeneity~  \\ 
\hline
MFT & 
Correct only when $U/t \lesssim 1$. & 
No gap/pseudogap in the normal state. &
No access to transport. &
Real space MFT is reasonable at $T=0$, can 
access size $\sim 40 \times 40$. 
\\
\hline
MFT+fluctuations & 
No results on the full $T_c(U)$ in the lattice model. & 
PG above mean field $T_{c}$, incapable of capturing BKT physics \cite{metzner}  &
No results & 
Not systematically explored.
\\ 
\hline 
SC-TMA  & 
Accurate upto intermediate coupling; captures non monotonic
behaviour correctly to $U\sim 8$ .\cite{sa-deisz}.   & 
Shows a PG, but quantitatively inaccurate due to neglect
of vertex corrections \cite{metzner}. & 
No results   &
Not generalised.
\\
\hline
2PSC & 
Fluctuations drive $T_{c}$ to zero \cite{metzner}. &
Accurate upto intermediate coupling \cite{sa-2psc}. &
No results  &
Not generalised.
\\ 
\hline 
XY models & 
Captures non monotonic $T_c(U)$ but not 
quantitatively accurate \cite{sa-kopec,sa-dupuis}. &
PG inferred from different pairing and $T_{c}$ scales. & 
Not explored.  &
Disordered XY model needs to be
derived from the Hubbard model.
\\ 
\hline
QMC & 
Accurate, sets the benchmark \cite{qmc-moreo1,qmc-paiv2}. &
Accurate in principle but involves uncertainties due to 
analytic continuation from imaginary frequencies \cite{qmc-moreo2,qmc-singer}. & 
Contains the relevant physics but dynamical
properties are difficult to extract due
to analytic continuation. &
Handles inhomogeneity \cite{bouad} but sizes
limited to $\sim 12 \times 12$.
Limited to   $U \gtrsim 4t$. 
\\
\hline
 DMFT &
Captures  non monotonic $T_c(U)$ but quantitatively inaccurate when
used in the 2D context \cite{dmft-keller}. &
Accurate \cite{dmft-dup}. &
Transport misses bosonic contribution.  &
Requires {\it ad hoc} real space generalisation.\\
\hline
SAF & 
Accurate, matches quantitatively with QMC.& 
Accurate, compares reasonably with QMC. &
Transport misses bosonic contribution. &
Handles inhomogeneity, ${\cal O}(N)$ method, readily
accesses size $\sim 30 \times 30$ \cite{trans}. 
\\ 
\hline
\end{tabular}
\caption{Comparison of earlier work with our method}
\end{center}

\end{table*}
%\newpage
%~~
%\newpage

\section{Model and Method}

We study the attractive two dimensional Hubbard model (A2DHM),
$$
H =
\sum_{ij,\sigma} (t_{ij} - \mu \delta_{ij})
  c_{i \sigma}^{\dagger} c_{j \sigma}
- \vert U \vert \sum_{i}
n_{i \uparrow} n_{i \downarrow}
$$
For us $t_{ij}$ denotes 
nearest neighbour tunneling amplitude $t$ on a
square lattice.  $\mu$ is the chemical potential.
We set $t=1$ and measure all other energies in terms of it.
$U >0$ is the strength of onsite Hubbard attraction.
We will focus on the density $n \sim 0.9$
which is close to half-filling 
but avoids the density wave features of $n=1$.

The model is known to have a superconducting ground 
state for all $n \neq 1$, while at
$n=1$ there is the coexistence of superconducting and  
density wave (DW) correlations in the ground state. 
For $ n \neq 1$ the ground state evolves from a
BCS state at $U/t \ll 1$ to a 
BEC of `molecular pairs' at $U/t \gg 1$.  
The pairing amplitude and gap at $T=0$ 
can be reasonably accessed within mean field theory or a 
simple variational wavefunction.

Mean field theory, however, assumes that the
electrons are subject to a {\it spatially uniform} 
self-consistent pairing amplitude $  \langle \langle 
c^{\dagger}_{i \uparrow} c^{\dagger}_{i \downarrow} \rangle \rangle$.
At small $U/t$ this vanishes when $k_B T \sim 
te^{-t/U}$, but at large 
$U/t$ it vanishes only when  $k_B T \sim U$. The 
actual $T_c$ at large $U$ is controlled by phase correlation 
of the local order parameter, rather than finite
pairing amplitude, 
and occurs at $k_B T_c \sim f(n)t^2/U$, where $f(n)$ is a 
function of the density. 
The wide temperature 
window, between the `pair formation' scale $k_B T_f
\sim U$ and $k_B T_c$ corresponds to equilibrium between
unpaired fermions and hardcore bosons (paired fermions). 

We use 
an auxiliary field scheme, based on the 
Hubbard-Stratonovich (HS) transformation, to
retain the thermal amplitude and phase fluctuations 
explicitly.
Formally, the HS transformation (below) allows us to recast
 the original interacting
problem into that of electrons coupled to space-time
fluctuating auxiliary fields. Usually
the auxiliary fields are {\it either} the
`pairing fields'  $\{\Delta_i(\tau),\Delta^*_i(\tau)\}$
{\it or} a `density' field $\phi_i(\tau)$.
The pairing field based decomposition is employed in the 
Bogoliubov-de Gennes (BdG) approach, while
QMC uses the density based decomposition.

While these two representations lead to the same answer when
the resulting problem is treated exactly, any approximation
leads to answers that are decomposition dependent.
`Single channel'
decompositions \cite{solms,dubi}, above, 
have the shortcoming that 
they {\it do not} reproduce the conserving 
Hartree-Fock-BdG (HFBdG) theory in their static limit,
crucial when density inhomogeneities are present.

Following recent suggestions \cite{conduit} 
we rewrite the A2DHM 
in terms of {\it classical}
auxiliary fields in {\it both} pairing and density
channels, see below, so that the saddle point reproduces HFBdG
theory.
\begin{eqnarray}
H_{eff} & = &
H_{kin}
+ \sum_{i} (\Delta_{i} c_{i \uparrow}^{ \dagger} c_{i \downarrow}^{ \dagger }
+ h.c.)  - \sum_{i} \phi_{i} n_{i} \cr
&& ~~~~~~+ \sum_{i} \frac{\vert \Delta_{i} \vert^{2}}{U}
+ \sum_{i} \frac{\phi_{i}^{2}}{U}
\end{eqnarray}
We treat the $\{\Delta_i,\Delta^*_i,\phi_i\}$
as time independent, {\it i.e} classical fields.
The partition function is:
$$
Z= \int {\cal D} \Delta {\cal D} \Delta^* {\cal D} \phi
Tr_{c,c^{\dagger}}
e^{-\beta H_{eff}}
$$

As $T \rightarrow 0$, it is obvious that $Z$ will be dominated by
the saddle point configuration, {\it i.e}, $\Delta_i$, $\phi_i$,
that minimise $H_{eff}$. Thus at $T=0$:
$ \Delta_i  = U \langle
c_{i \uparrow} c_{i \downarrow}
\rangle$,
$ \phi_i = {U \over 2} \langle  n_i  \rangle $.
These minimum conditions are easily recognised as the
self-consistency requirements of HFBdG theory.
Minimising $H_{eff}$ is thus the same as solving the 
Bogoliubov-de Gennes (BdG) equations.
While this is good, we 
want to capture the thermal physics as well.

At finite $T$
the BdG approach would promote the consistency condition
above to thermal averages, and the electrons would see
only the  {\it average} background at any temperature.
It is easy to see the limitation of this approach when
we consider the clean problem at large $U$. Since the
system is translation invariant, averages like
$\langle
c_{i \uparrow}^{ \dagger} c_{i \downarrow}^{ \dagger}
\rangle
$ are site independent. Their self-consistent value,
and hence superconductivity, vanishes when $T \sim U$,
way above the correct $T_c \sim t^2/U$.
To remedy this we need to  {\it average the electron motion
over disordered thermal configurations}
 rather than
solve for electron motion in the disorder averaged
configuration.

The Boltzmann weight for the occurence of a particular
$\{\Delta_i,\phi_i\}$ configuration is
\begin{equation}
P\{\Delta_i,\phi_i\} \propto Tr_{c,c^{\dagger}}
e^{-\beta H_{eff}}
\end{equation}
This is related to the electron free energy in a
particular
$\{\Delta_i,\phi_i\}$ background. If the fields are
large and random the trace cannot be analytically computed.
We generate the equilibrium $\{\Delta_i,\phi_i\}$
configurations by a Monte Carlo technique \cite{dubi,dag}.
This involves diagonalisation of the electron Hamiltonian
$H_{eff}$ for every attempted update of the auxiliary fields
and the Metropolis condition for acceptance of this move.
We use a cluster based implementation  of the
Monte Carlo \cite{tca} and can access sizes upto $32 \times 32$.

The expectation value of any observable $f(c_{i},c^{\dagger}_{i})$ 
can then be computed as follows. For each configuration 
$\{\Delta_i,\phi_i\}$, the first step is to 
express the fermionic operators in terms 
of the quasiparticle operators $(\gamma_{i},\gamma^{\dagger}_{i})$,
with the HFBdG eigenvectors $(u^{i}_{n}, v^{i}_{n})$ as coefficients.
The electronic trace over the resultant expressions 
in terms of the quasiparticle operators can be easily evaluated. 
This final expression is then averaged over
all thermal configurations $\{\Delta_i,\phi_i\}$.
We will describe the indicators separately
in detail in their respective sections.

\section{Results}

\subsection{Thermodynamic indicators}

%------------------------------------------------------------------------
\begin{figure}[b]
\centerline{
~
\includegraphics[width=3.9cm,height=3.6cm,angle=0,clip=true]{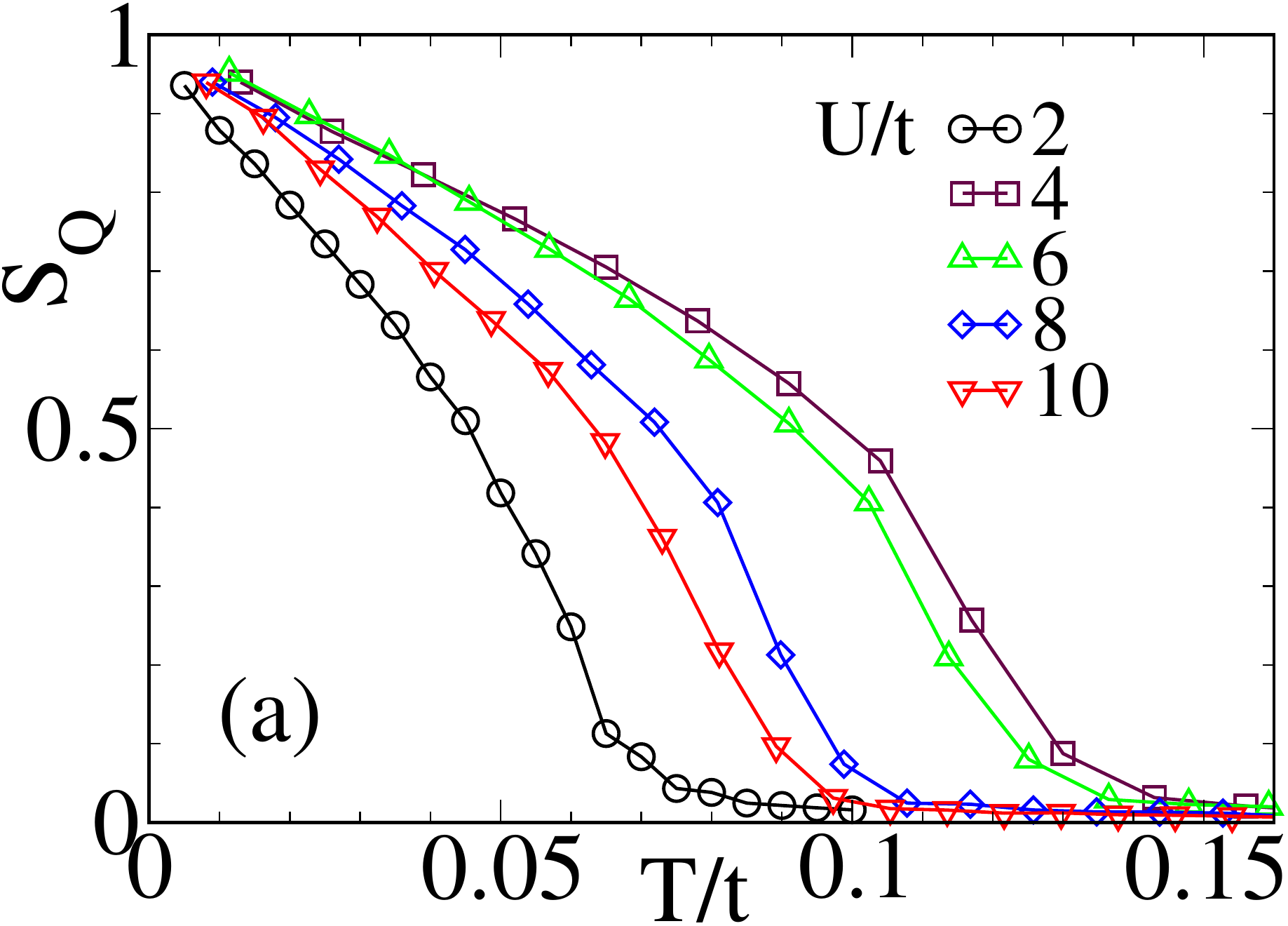}
\hspace{.051cm}
\includegraphics[width=4.0cm,height=3.6cm,angle=0,clip=true]{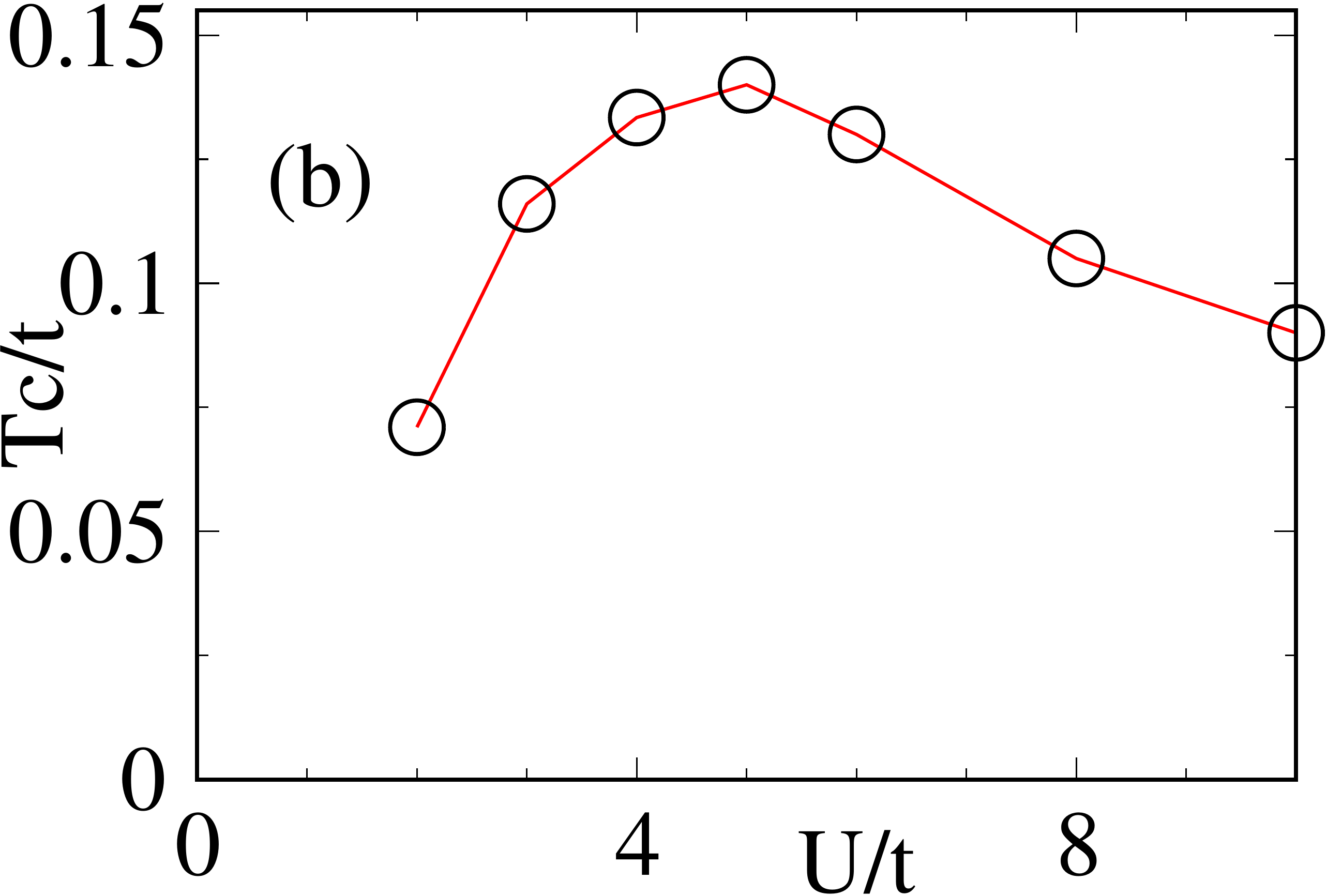}
}
\vspace{.2cm}
\centerline{
\includegraphics[width=4.1cm,height=3.6cm,angle=0,clip=true]{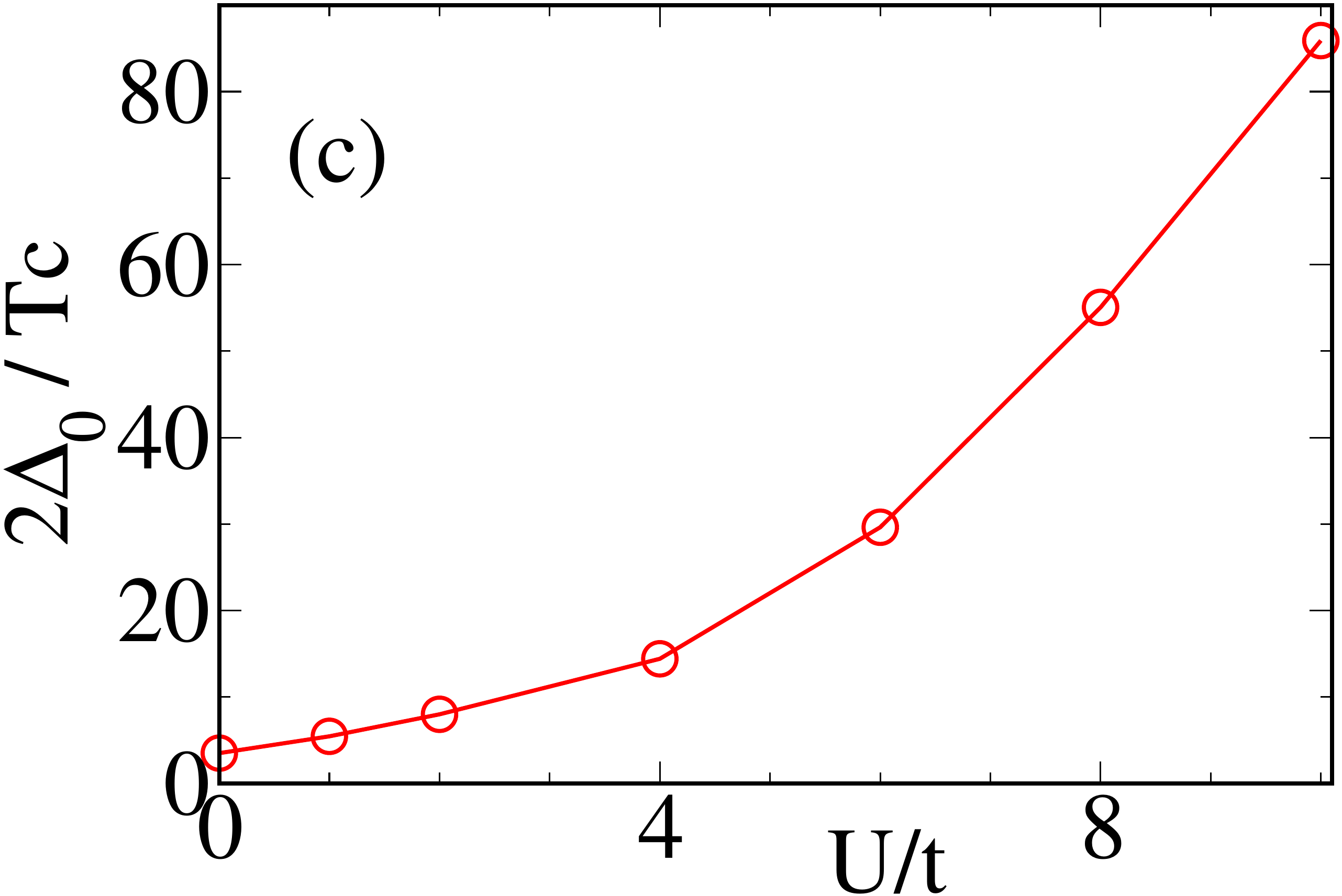}
\hspace{.2cm}
\includegraphics[width=3.8cm,height=3.6cm,angle=0,clip=true]{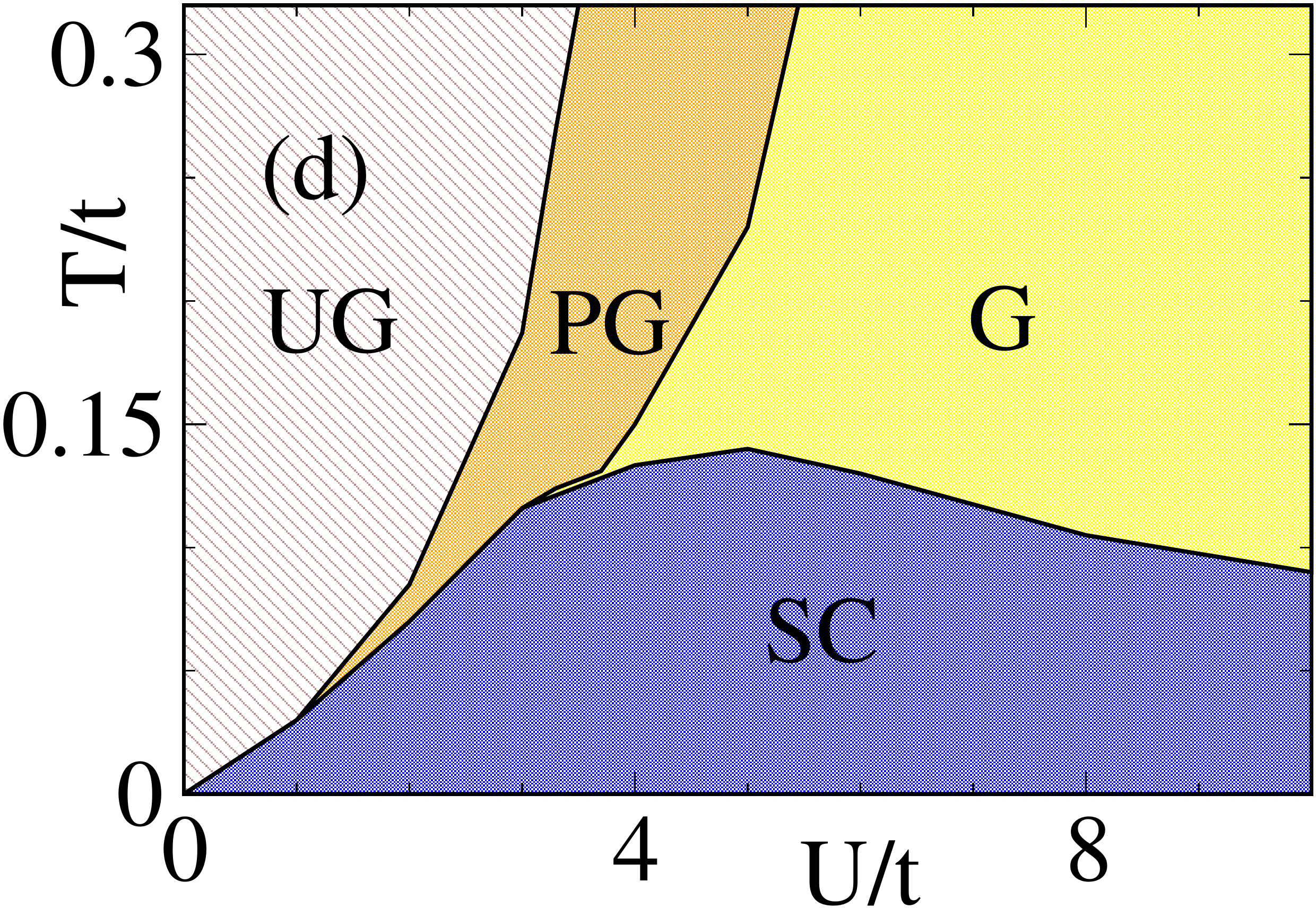}
}
\caption{Colour online: (a).~Temperature dependence of 
the ${\bf q}=\{0,0\}$ component of the
pairing field correlation for different $U/t$. 
The onset locates the superconducting $T_c$. (b).~The $T_c$
inferred from the structure factor result.
This is compared to QMC results at the end of the paper.
(c).~Ratio of $T=0$ gap $2 \Delta_0$ to $T_c$.  
In the BCS limit the ratio would be $\sim 3.5$. (d).~The `phase
diagram' in terms of the low frequency behaviour of the density
of states. The high temperature normal state has three regimes,
ungapped (UG), pseudogapped (PG) and gapped (G), while for $T < T_c$
the system is a gapped superconductor (SC).
}
\end{figure}
%------------------------------------------------------------------------

Fig.1.(a)  shows our result for the structure factor
corresponding to the growth of superconducting order.
We compute the thermally averaged 
pairing field correlation
$S({\bf q}) = {1 \over N^2} \sum_{ij} 
\langle \Delta_i   \Delta^{\star}_j \rangle
e^{i {\bf q}.{(\bf r}_i -{\bf r}_j)} $ 
at ${\bf q} = \{0,0\}$. This is like the 
`ferromagnetic' correlation
between the $\Delta_i$, treating them as two dimensional moments. If
the ${\bf q} = \{0,0\}$ component, $S(0)$, is ${\cal O}(1)$ it implies
that the pairing field has a non-zero spatial average and would in
turn induce long range order (power law correlation in 2D)
in the thermal and quantum 
averaged correlation $ M_{ij} =
\langle \langle c^{\dagger}_{i\uparrow} c^{\dagger}_{i\downarrow}
c_{j\downarrow} c_{j\uparrow} \rangle \rangle$. 
We locate the superconducting
transition from the rise in $S(0,T)$ as the system is cooled. 
The results are not reliable
below $U/t \lesssim 1$, since the correlation
length $\xi$ 
becomes comparable to our system size, but 
compare very well with available QMC data for $U/t \gtrsim 2$.
We will discuss the interpretation of the $S(0,T)$ results
in detail at the end of the paper.

Panel 1.(b) shows the result for $T_c(U)$ showing the clear peak
around $U/t \sim 5$. We will compare this to the result from QMC,
and also discuss the system size dependence,
at the end of the paper.

% --------------------------------------------------
\begin{figure}[t]
\centerline{
\includegraphics[width=8.4cm,height=5.2cm,angle=0,clip=true]{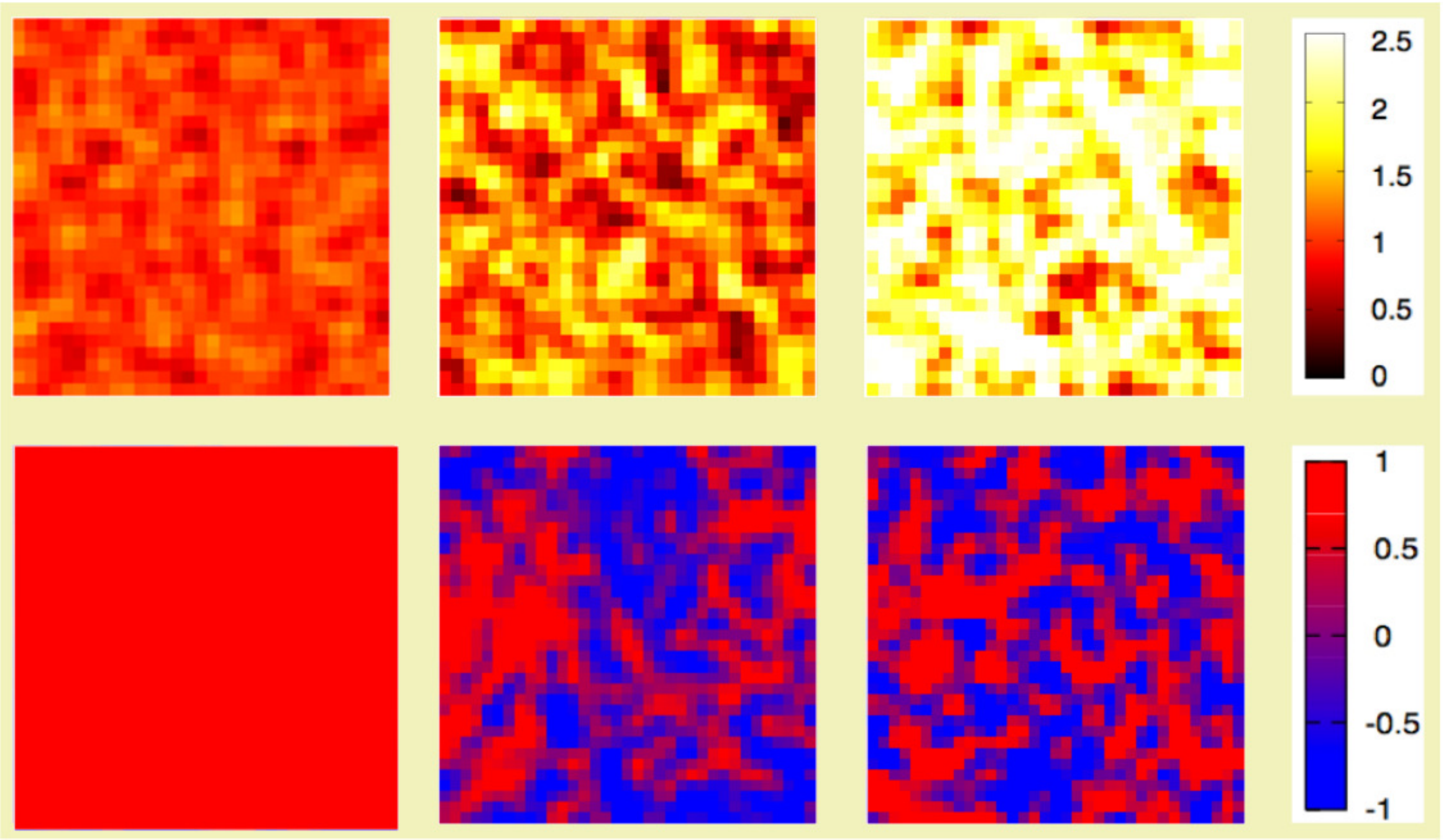}
}
\vspace{0.3cm}
\centerline{
\includegraphics[width=8.4cm,height=5.2cm,angle=0,clip=true]{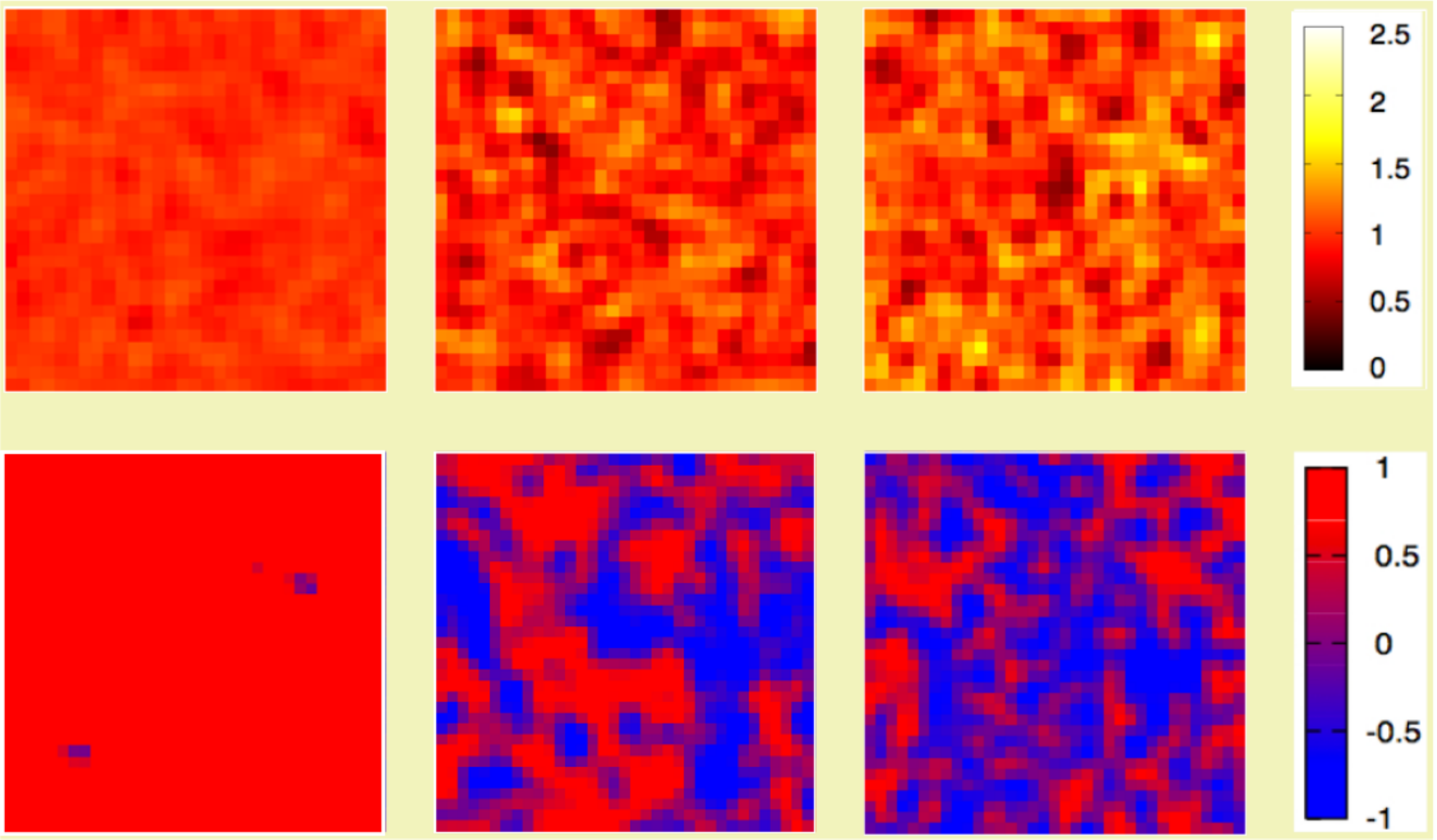}
}
\vspace{0.3cm}
\centerline{
\includegraphics[width=8.4cm,height=5.2cm,angle=0,clip=true]{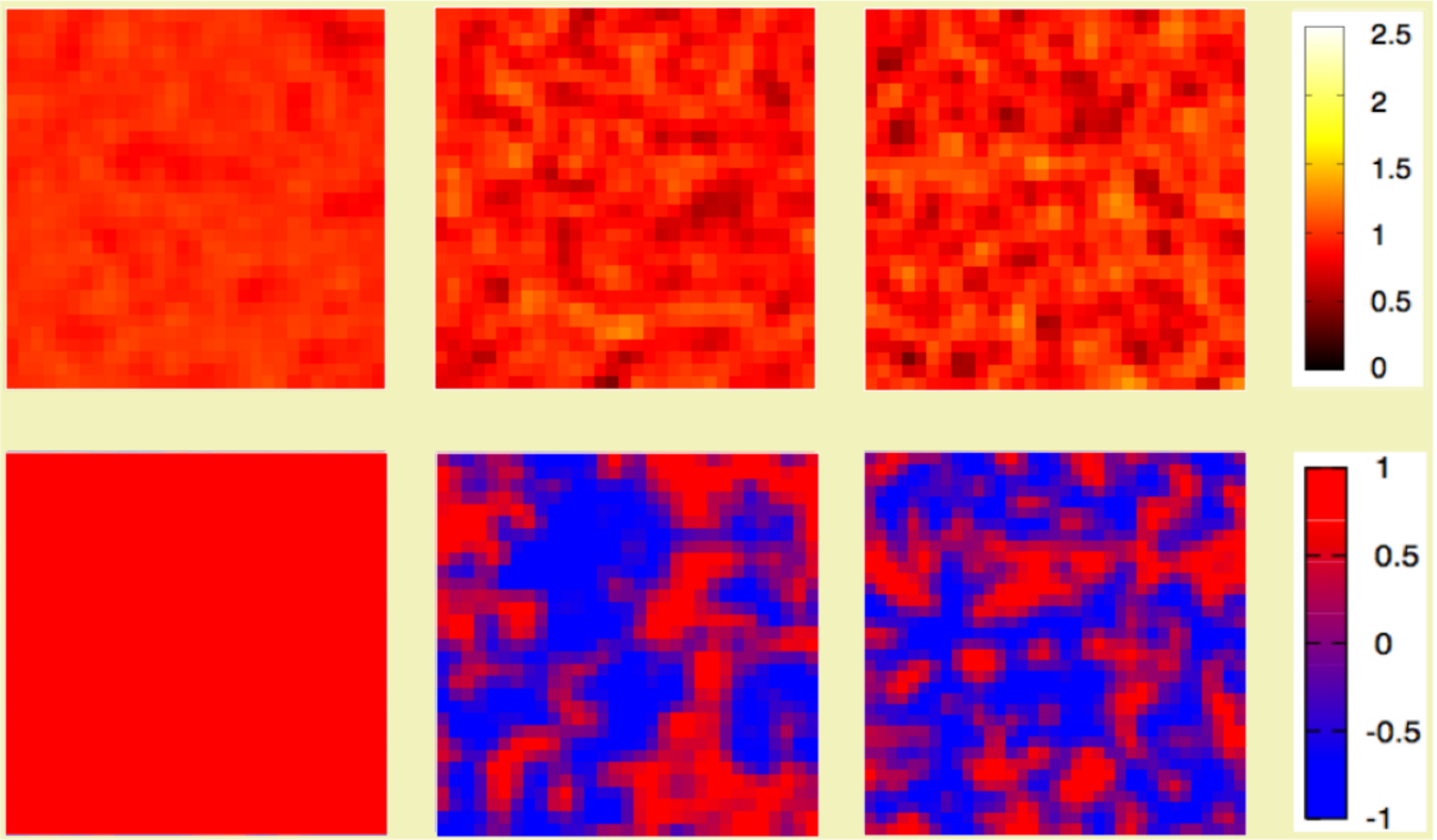}
}
\caption{Colour online: Maps of amplitude fluctuation and 
phase correlation for
single configurations
at $U/t=2$ (top), $U/t=6$ (middle) and $U/t=10$ (bottom) 
at three temperatures: 
$T=0.1T_c$, $T=T_{c}$ and $T=2T_{c}$ (left to right). For
each set, the upper row 
shows the amplitude $\vert \Delta_i \vert$ (normalised by the $T=0$ 
mean field value $\Delta_0$) for a MC configuration, while the
lower row shows the phase correlation:
$\Phi_i = cos(\theta_i - \theta_0)$, where $\theta_0$ is the
phase at a site ${\bf R}_0$  near the center. 
}
\end{figure}
% --------------------------------------------------

Panel 1.(c)  highlights the rapid rise in the `gap'
to $T_c$ ratio with increasing interaction. In the weak coupling
limit this value is $3.5$, at $U=2t$ it is already $\sim 7$,
quite beyond BCS, and grows roughly as $(U/t)^2$ at large $U$.
Needless to say, the $T=0$ gap is not an indicator of the robustness
of the superconducting state once we go beyond weak coupling.

The large gap but low $T_c$ leaves its imprint on several 
physical properties.
The phase diagram, Fig.1.(d), highlights this.
At weak coupling the vanishing of SC order also means the
vanishing of the gap in the density of states.
The high $T$ regime at small $U/t$ is ungapped (UG).
 The regime $U/t \lesssim 2$
is a `renormalised BCS' window, although the gap to $T_c$ ratio
is large. For $2 \lesssim U/t \lesssim 4$ the $T > T_c$ phase
has a pseudogap (PG), which we show in Fig.2, while for 
$U/t \gtrsim 4$
the $T > T_c$ regime is gapped. Notice that the normal state
gap appears before the peak $T_c$ is reached, {\it i.e}, on
the `BCS' side of the crossover.

\subsection{Background fields}

%------------------------------------------------------------------------
\begin{figure}[t]
\centerline{
\includegraphics[width=4.0cm,height=3.9cm,angle=0]{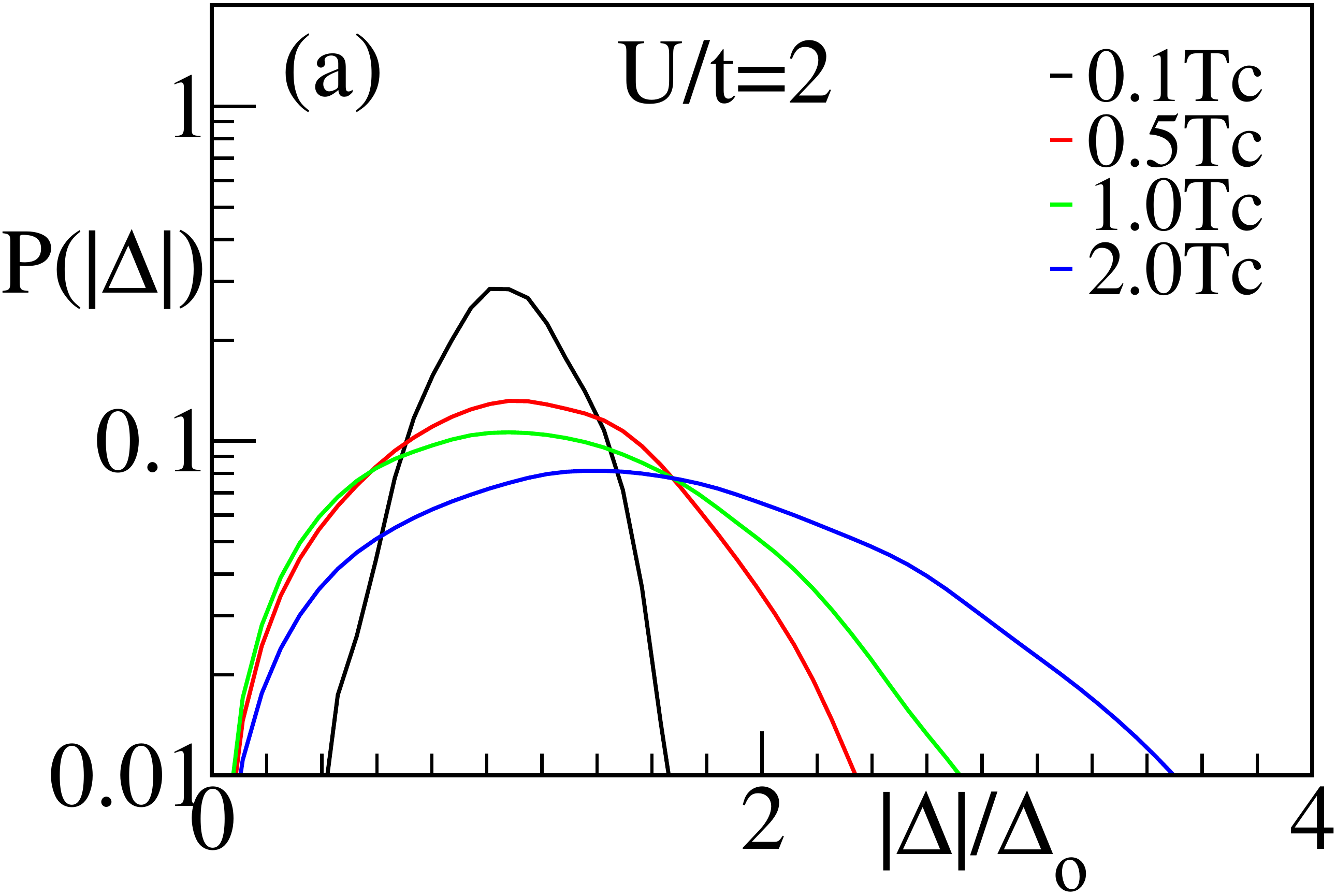}
%\hspace{.3cm}
\includegraphics[width=4.0cm,height=3.9cm,angle=0]{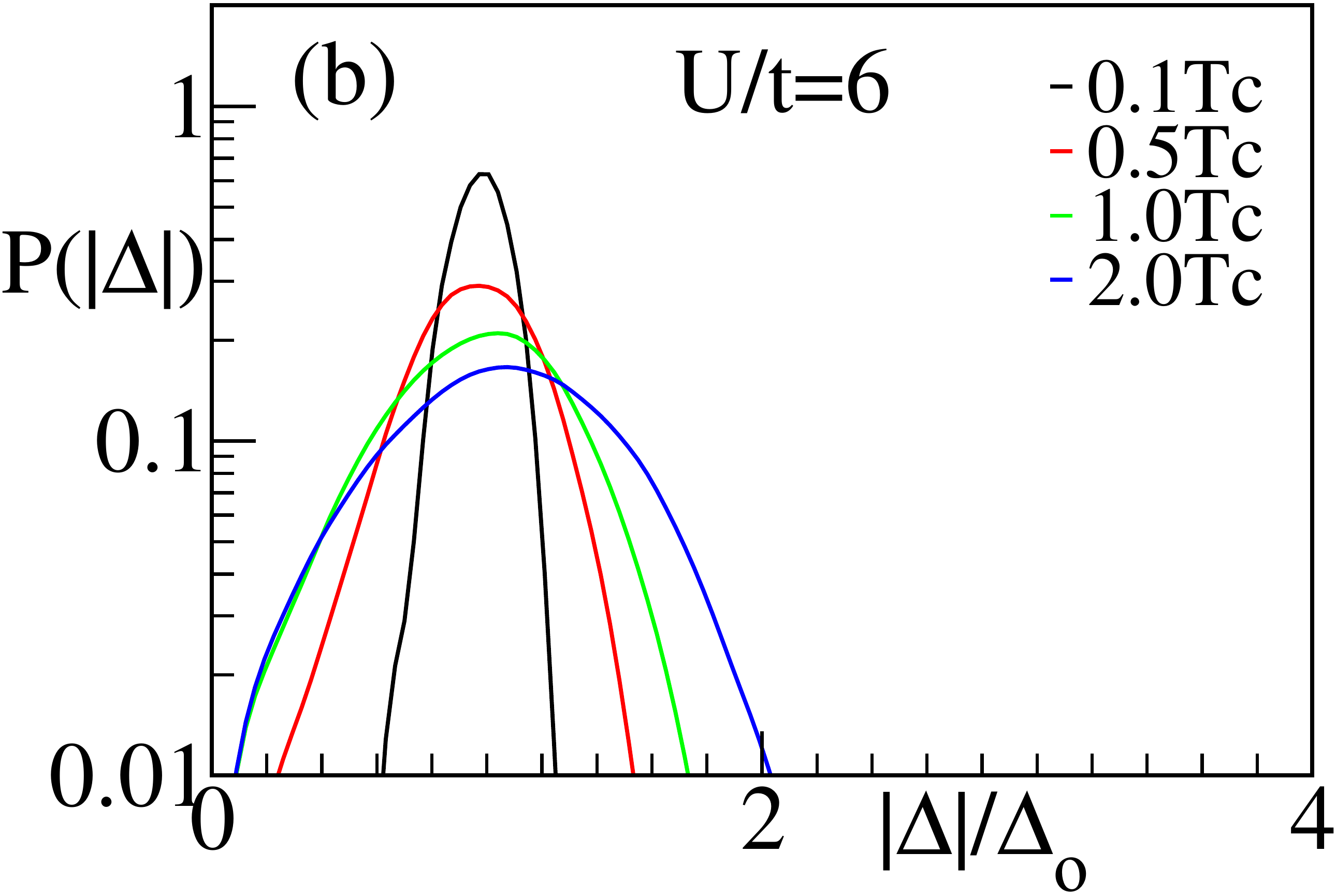}
}
\vspace{.2cm}
\centerline{
\includegraphics[width=4.0cm,height=3.9cm,angle=0]{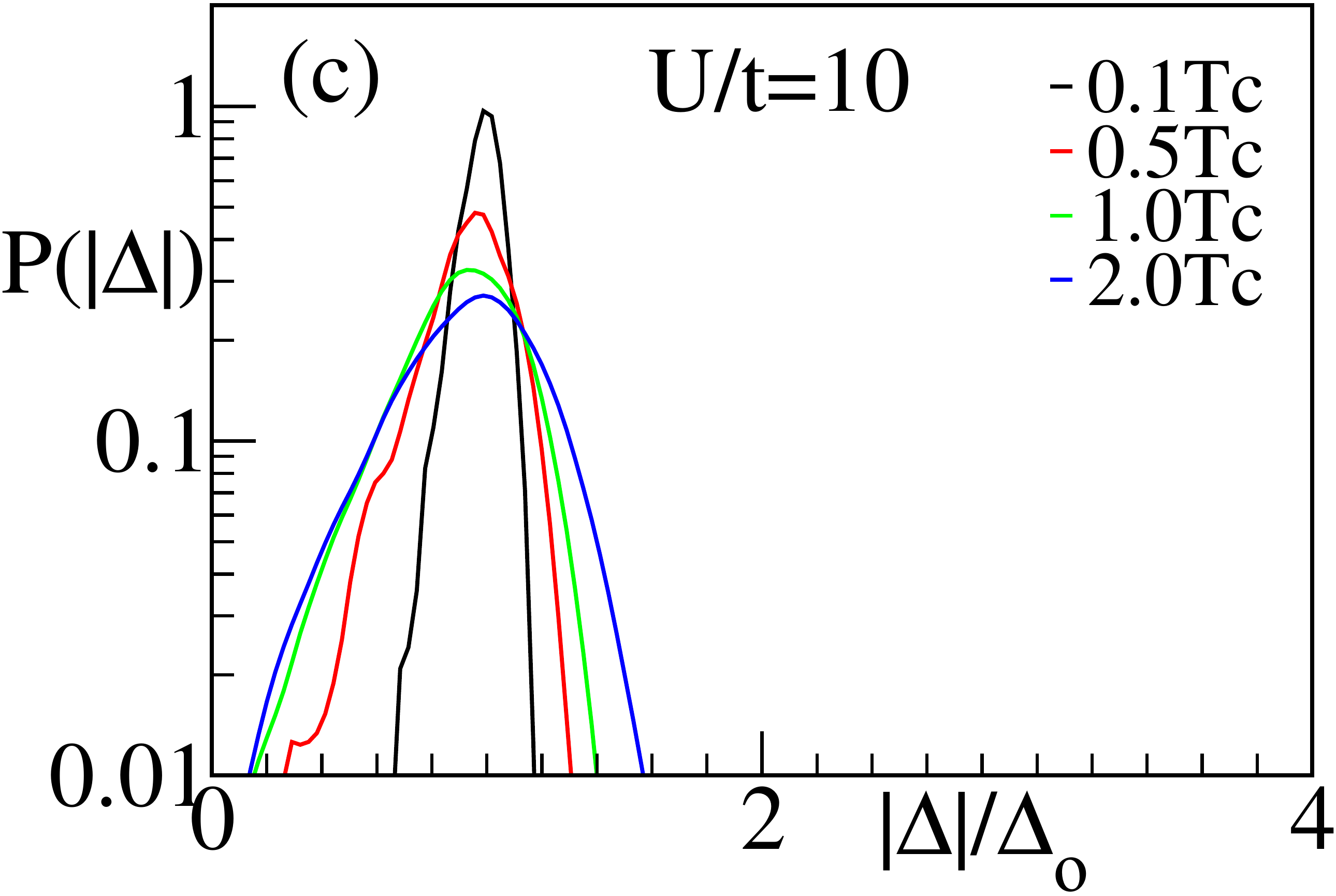}
%\hspace{.3cm}
\includegraphics[width=4.1cm,height=3.9cm,angle=0]{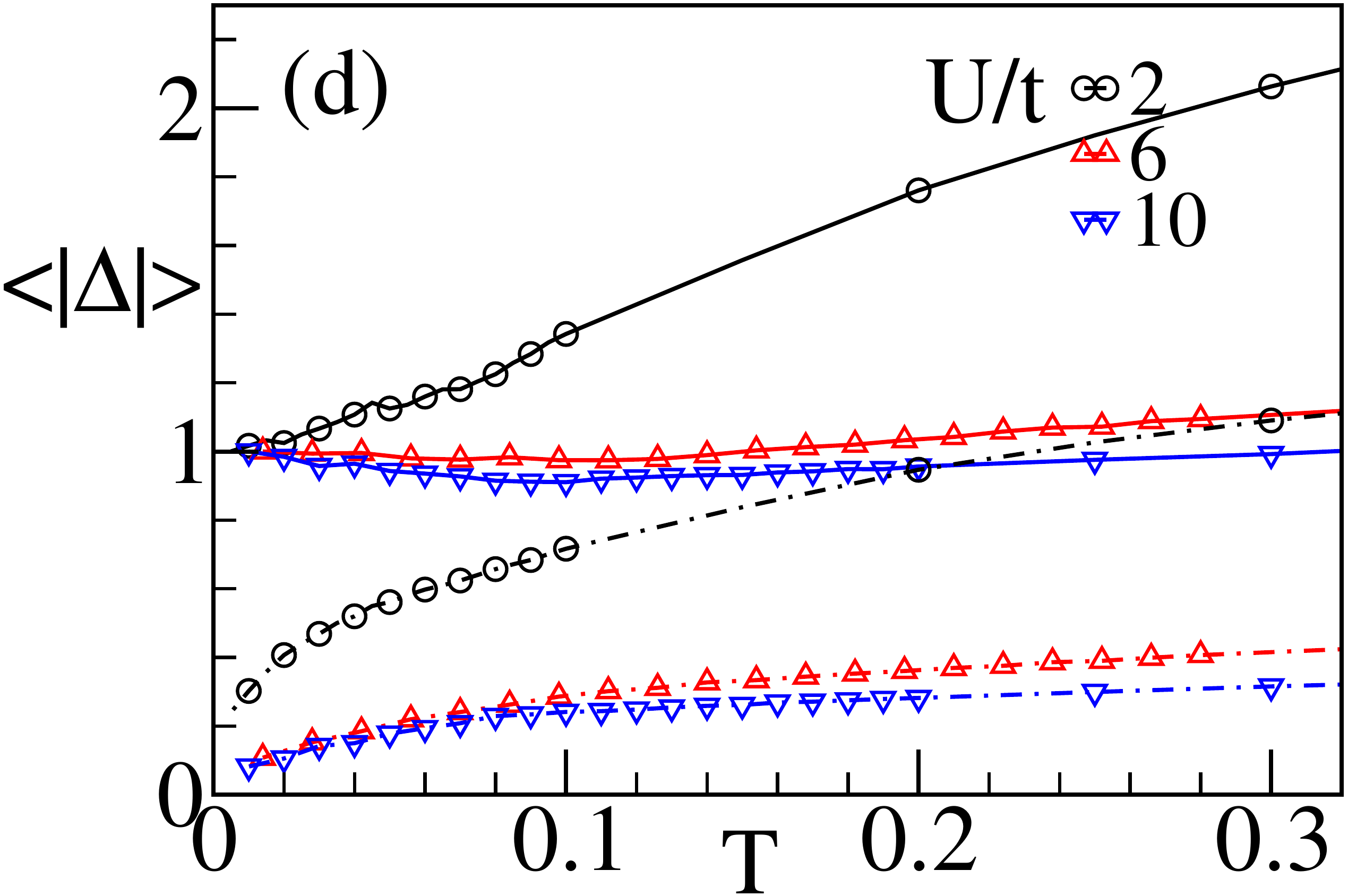}
} 
\caption{Colour online: (a)-(c).~The distribution 
$P(\vert \Delta \vert)$ of the 
magnitude, $\vert \Delta \vert$,
of the pairing field. The x-axis is normalised
by the mean field value 
$\Delta_0 $ at $T=0$.
The results are for $T = 0.1T_{c},~0.5T_{c},~1.0T_{c},~2.0T_{c}$.
(a).~$U/t=2$, (b).~$U/t=6$, (c).~$U/t=10$.
At $U/t=2$ there is a prominent increase in the mean
and width of $P(\vert \Delta \vert)$ with $T$. This $T$ 
dependence weakens with growing $U/t$.
(d).~The growth of the mean value and width with $T$. The mean
is normalised by the $T=0$ value. The firm
lines denote the mean $|\Delta|$, while the dot-dashed lines show the 
corresponding width.
}
\end{figure}
%------------------------------------------------------------------------

To understand the spatial behaviour of the system and its evolution
with $U$ and $T$ we examine the variation of the background
fields  $\Delta_i$ and $\theta_i$.
Fig.2 shows single snapshots of $\vert \Delta_i \vert$
(upper row in each set), normalised
by the $T=0$ mean field value, and 
the phase correlation 
$\Phi_i = cos(\theta_i - \theta_0) $
(lower row in each set),
where $\theta_0$ is the angle at fixed
site $R_0$ in the lattice. 

Of the three sets in Fig.2, the top set is
for  $U/t =2$, which we will use as
typical of `weak' coupling, the middle set is
for $U/t=6$, typical of intermediate coupling,
and the top set is for $U/t=10$, strong coupling.
The rows are for $T/T_c(U)=0.1,~1,~2$.

At $T=0$
all the snapshots show almost 
uniform $ \vert \Delta_i 
\vert$ and perfect phase locking at all $U/t$.
This is just the mean field state.
As we move to higher $T$, however, we see 
a clear difference in the amplitude
fluctuations of the three systems. While the 
$U/t=2$ plots show an increasing 
inhomogeneity and a steadily rising value 
of $\Delta$ throughout the system,
the $U=10t$ case hardly shows any change. The $U=6t$ 
behaviour is intermediate.
This shows that with increasing $U$, the system moves 
smoothly from an amplitude fluctuation
dominated regime to one in which amplitudes are 
effectively constant, the transition
being driven by the phase fluctuations.

The phase maps, on the other hand, show how the 
system breaks up into correlated patches
with temperature. The middle column corresponds 
to $T_{c}$, and show large correlated 
clusters, as expected for a system close to 
criticality. As $T$ is increased, the 
correlation length decreases, as evident from the 
right column at $2T_{c}$. 

While it is phase fluctuations that ultimately
destroy order at all $U/t$, the amplitude fluctuations
are quantitatively important at 
weak coupling.
To highlight this,
we plot the distribution of $| \Delta |$ for the 
three $U$ values at four temperatures
$T=0.1T_{c}$, $0.5T_{c}$, $T_{c}$ 
and $2T_{c}$ for each $U$ in Fig.3 (a),(b) and (c). (d) shows the
temperature dependence of the mean $\langle |\Delta| \rangle$
and its variance for $U=2t,~6t$ and $10t$.
We find that the distributions widen for each case 
with increasing $T$, but the increase
is much more pronounced at weak coupling, and 
decreases systematically with increasing
coupling. 
The distribution is noticeably non gaussian at high
temperature in the weak coupling case.
The temperature dependence of the mean and width of 
$P(\vert \Delta \vert)$ is shown in Fig.3.(d). A detailed
discussion is postponed to the end of the paper.

% We will relate the results here to the more 
% traditional issue of fluctuation of the `order
% parameter', ${\tilde \Delta_0} = {\tilde \Delta_{{\bf q}=0}} $,
% where 
% $$
% {\tilde \Delta_{\bf q}} = (1/N) \sum_i \Delta_i e^{i {\bf q}.{\bf R}_i}
% $$
% in the discussion section.
% That would focus on the 
% zero momentum component of the pairing field,
% as in usual Ginzburg-Landau theories.
\subsection{Density of states}

% ---------------------------------------------
\begin{figure}[]
\centerline{
\includegraphics[width=4.1cm,height=3.0cm,angle=0]{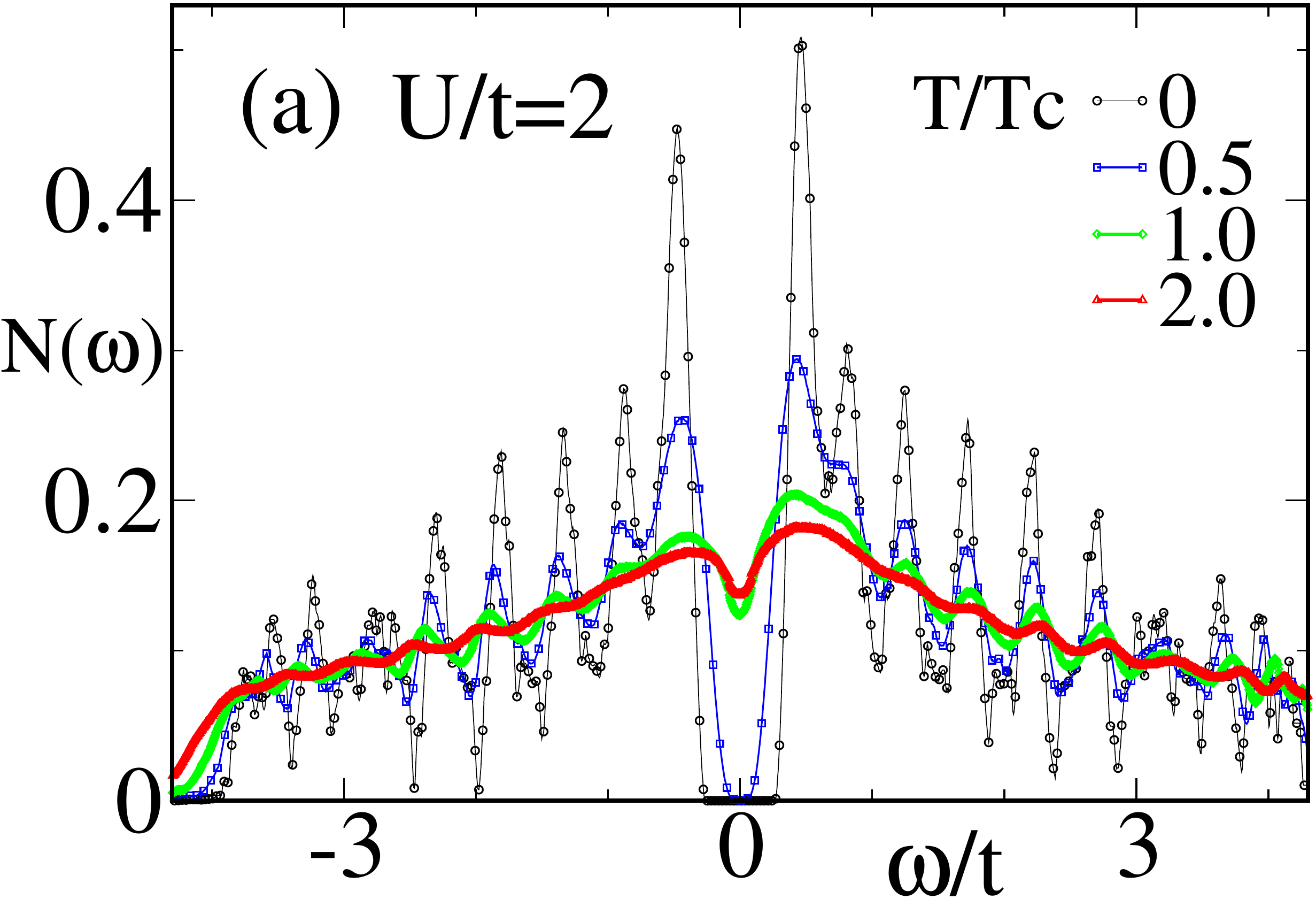}
\hspace{.1cm}
\includegraphics[width=4.1cm,height=3.0cm,angle=0]{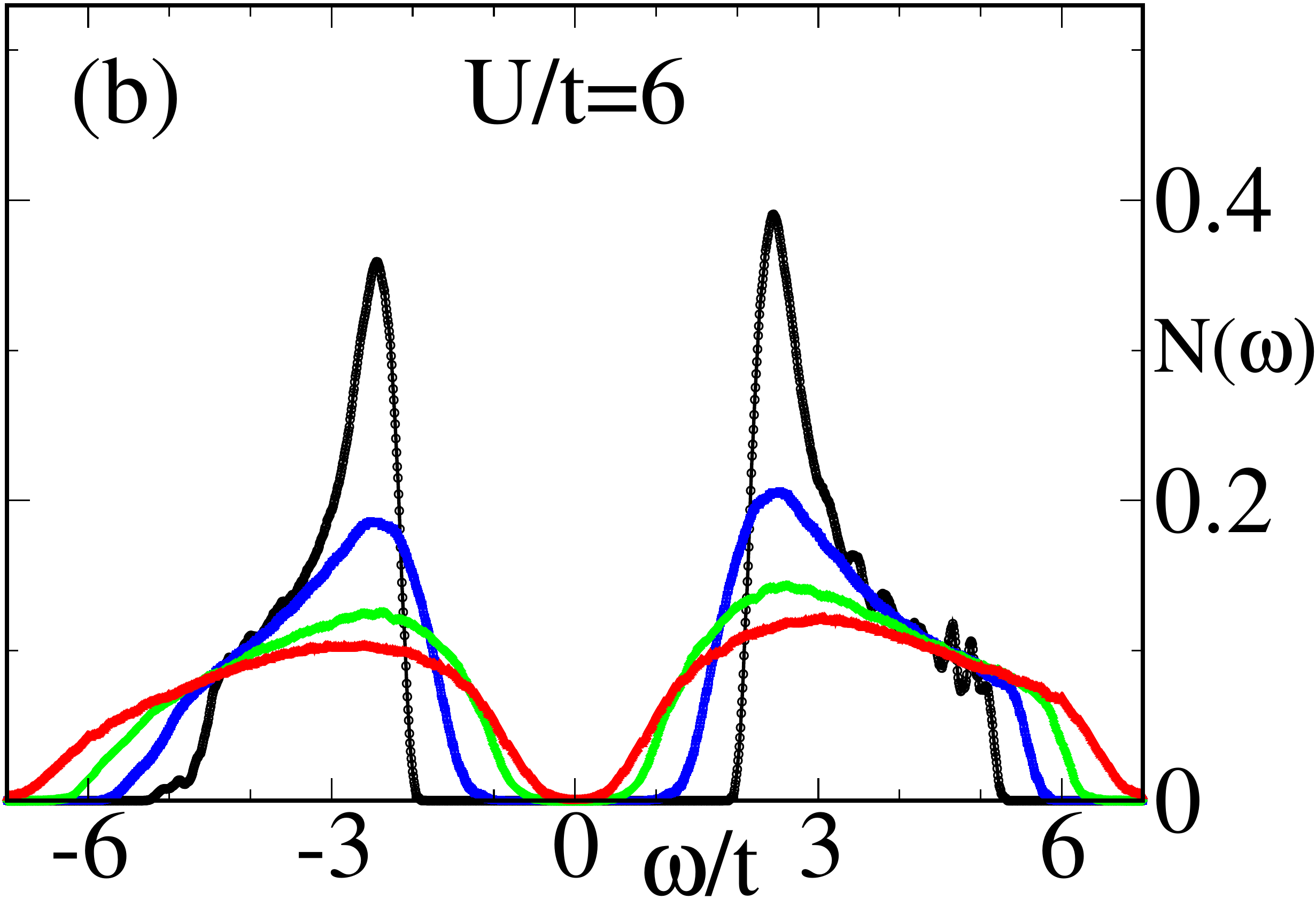}
}
\vspace{.1cm}
\centerline{
\includegraphics[width=4.1cm,height=3.0cm,angle=0]{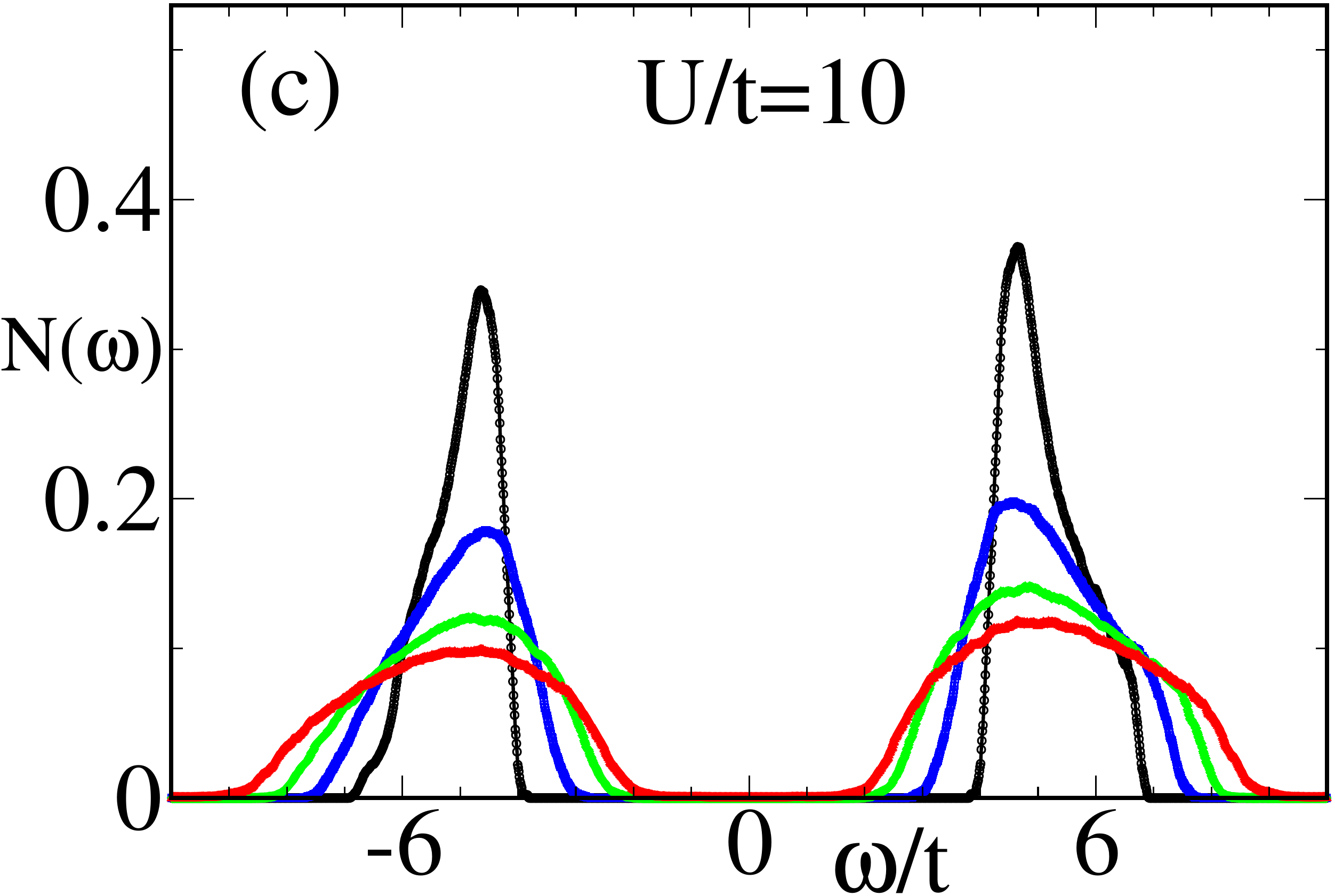}
\hspace{.1cm}
\includegraphics[width=3.9cm,height=3.0cm,angle=0]{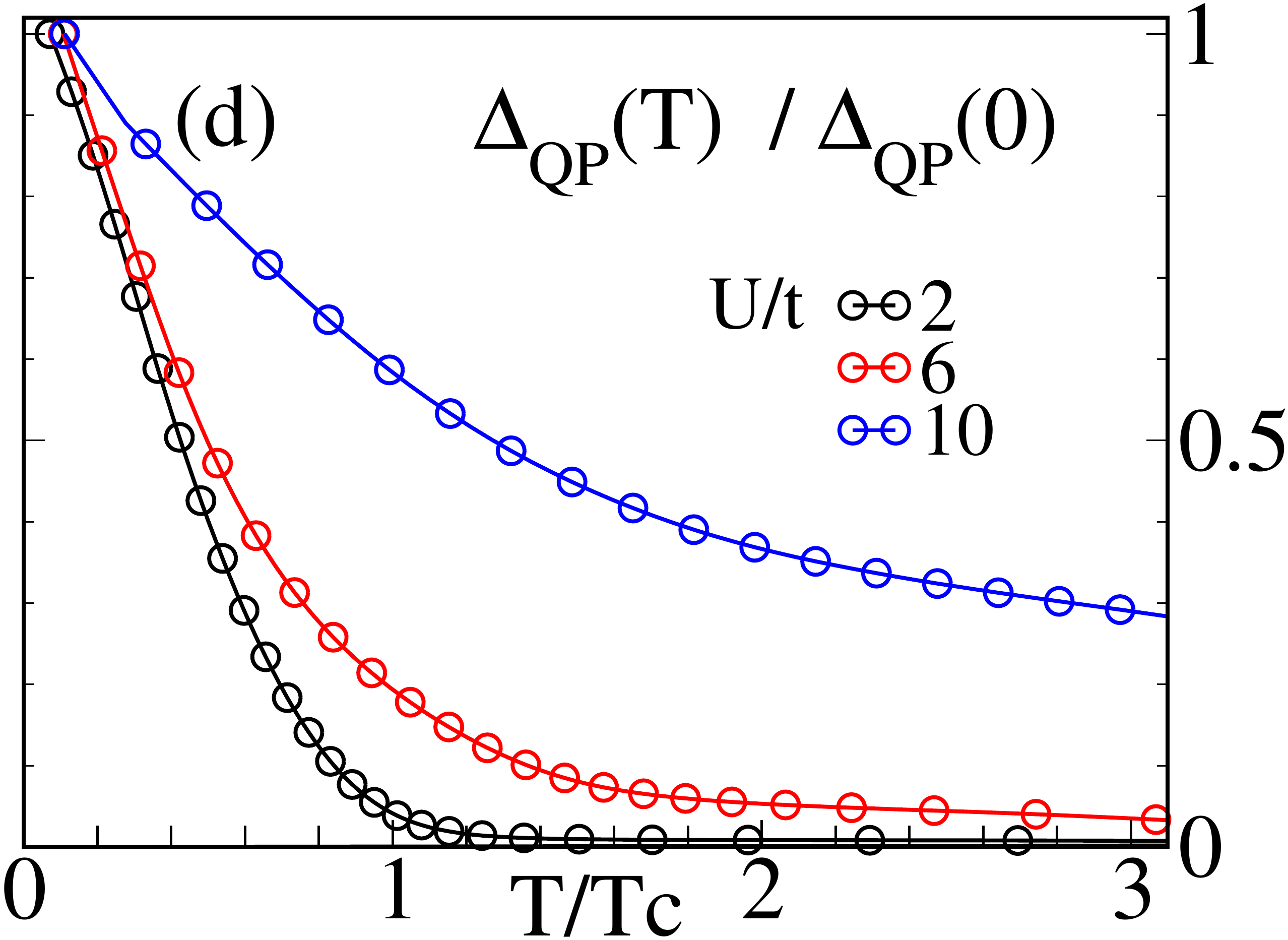}
~
}
\caption{Colour online: Temperature dependence of the
DOS, $N(\omega)$ at different couplings.
Panels (a)-(c) have the same legends.
(a).~$U/t=2$, (b).~$U/t=6$, and (c).~$U/t=10$. 
The oscillations in the DOS
in panel (a) are finite size artifacts (even on a $24 \times 24$
lattice). At $U/t=2$ the gap essentially vanishes at $T \sim T_c$,
while at $U/t=6$ a small `hard gap' persists to  $T_c$ and
above, although lorentzian broadening gives the impression of
a pseudogap at the highest $T$.
For $U/t=10$ a `hard gap' persists to
$T \sim 0.5$ although with a clear
reduction with increasing temperature.
(d).~Variation in the single
particle gap, normalised
by its $T=0$ value.
}
\end{figure}
% ---------------------------------------------

Fig.4 highlights the behaviour of 
the single particle density of states
(DOS). We show results at the `BCS end' $(U=2t)$, near the peak
$T_c$ $(U=6t)$, and in the BEC regime $(U=10t)$.
The $T=0$ results in all cases are described by the
canonical DOS, $N(\omega) \sim 1/\sqrt{ \omega^2 - 
\vert \Delta_0 \vert^2}$, where $2\Delta_0$ is the full
$T=0$ gap in the single particle spectrum. There is a gap
in the spectrum at all $U$, a `coherence peak' at the gap
edges, arising from electron propagation in a perfectly
pair correlated background, and a featureless fall at 
high energies. The oscillatory pattern in the DOS at 
$T=0$ for $U=2t$ is a consequence of finite size,  showing 
up even on a $24 \times 24$ lattice.

While the $T=0$ DOS is just a mean field result, and the
$T$ dependence at $U=2t$ is expected,
the $T$ dependence at $U=6t$ and $U=10t$ is not 
obvious. At $U=6t$, the system has a `hard gap'
persisting to $T \sim 3T_c \sim 0.5t$.
This is reflected in the phase
diagram in Fig.1. 
At both $U=6t$ and $10t$ the transition
to SC occurs from a gapped fermion state 
rather than a Fermi liquid. 
The DOS indicates that down to
$U$ values around peak $T_c$ (and even somewhat below) 
the qualitative physics remains similar to
the BEC end.

Another striking feature is the large transfer of spectral
weight that occurs on a modest change of temperature. For
the $U=10t$ case, for example, at $T \sim T_c \sim 0.09t$ 
there is weight transfer over a scale ${\cal O}(U)$. 
The reason is fairly simple: the magnitude 
$\vert \Delta_i \vert$
in this limit are almost $T$ independent, but the phase
correlation between them is destroyed at a temperature
$T \sim t^2/U$. As a result, over a small $T$ window
the
system evolves from a state with perfectly ordered
$\Delta_i$, to one where these large amplitudes are
randomly oriented.
The strong `disorder' in the $\Delta_i$ lead to the
broadening of the density of states.

 The large size of $\vert \Delta_i \vert$
even in the normal  
state preserves the gap feature, but the randomness smears
the band edges.

\subsection{Spectral functions}

%------------------------------------------------------------------------
\begin{figure}[b]
\centerline{
\includegraphics[width=9.0cm,height=10.0cm,angle=0]{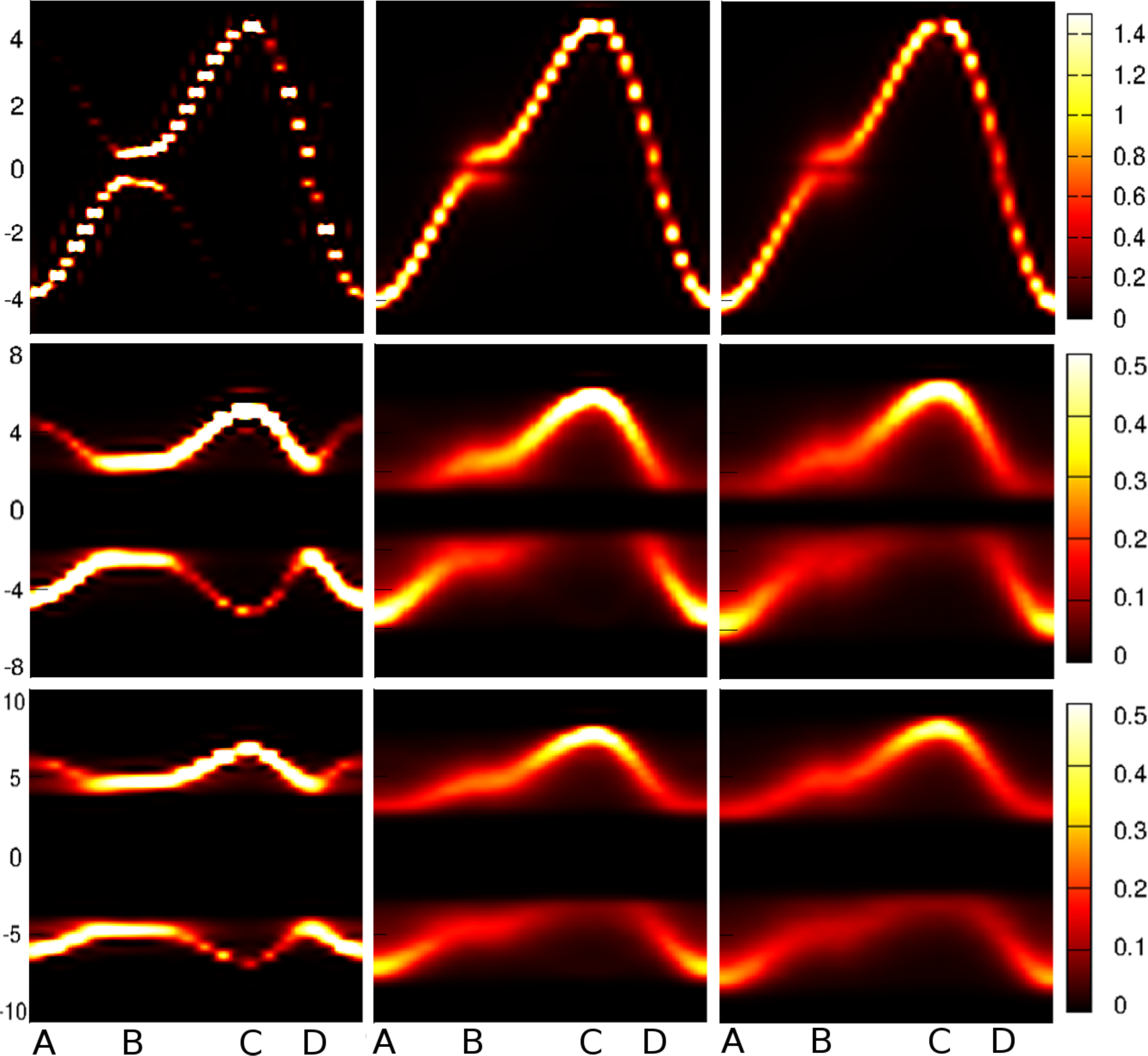}
}
\caption{Colour online: 
Plot of $A({\bf k}, \omega)$. The rows, left to right, are for
$U/t=2$, $6$ and $10$. Columns, top to bottom,
correspond to $0.1T_{c}$,$T_{c}$ and $2T_{c}$.
The momentum, on the x-axis, is scanned as
($0$,$0$) $\rightarrow$ ($0$,$\pi$)
$\rightarrow$ ($\pi$,$\pi$) and back through ($\pi/2$,$\pi/2$) to ($0$,$0$) 
along the diagonal. 
These points are labelled as A, B, C and D respectively.
The gaps are lowest around ($\pi/2, \pi/2$) 
and ($\pi, 0$), where the Fermi-surface of
the free system intersects our path in k-space. Increasing temperature causes 
broadening and a decrease of the gaps, which close in the case of $U=2t$. 
The increasing
symmetry of the low $T$ graphs with increasing $U$ signals the participation 
of  states far from the FS in pairing.
}
\end{figure}
%------------------------------------------------------------------------

%------------------------------------------------------------------------
\begin{figure*}[t]
\centerline{
\includegraphics[width=5.5cm,height=3.9cm,angle=0]{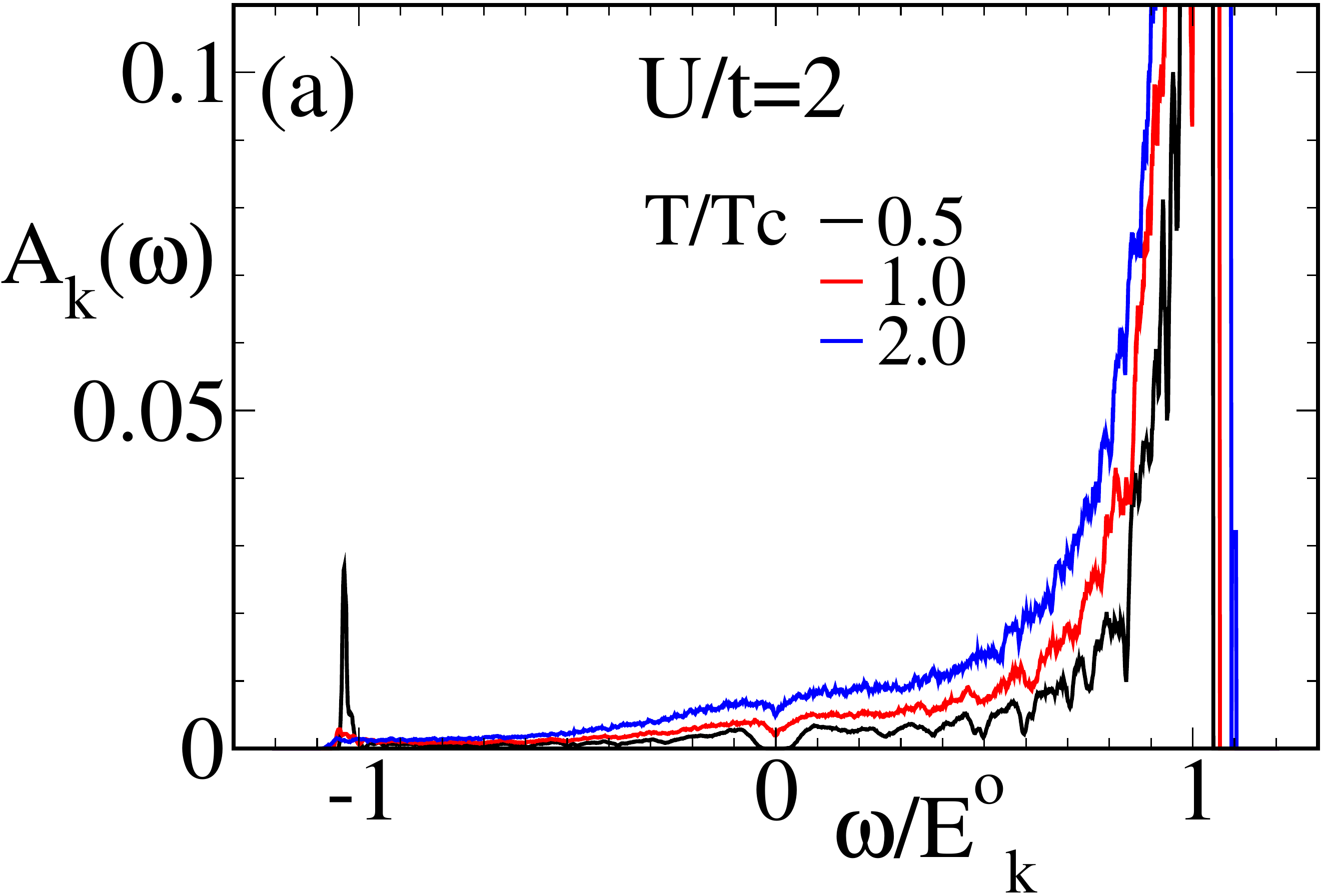}
\includegraphics[width=5.5cm,height=3.9cm,angle=0]{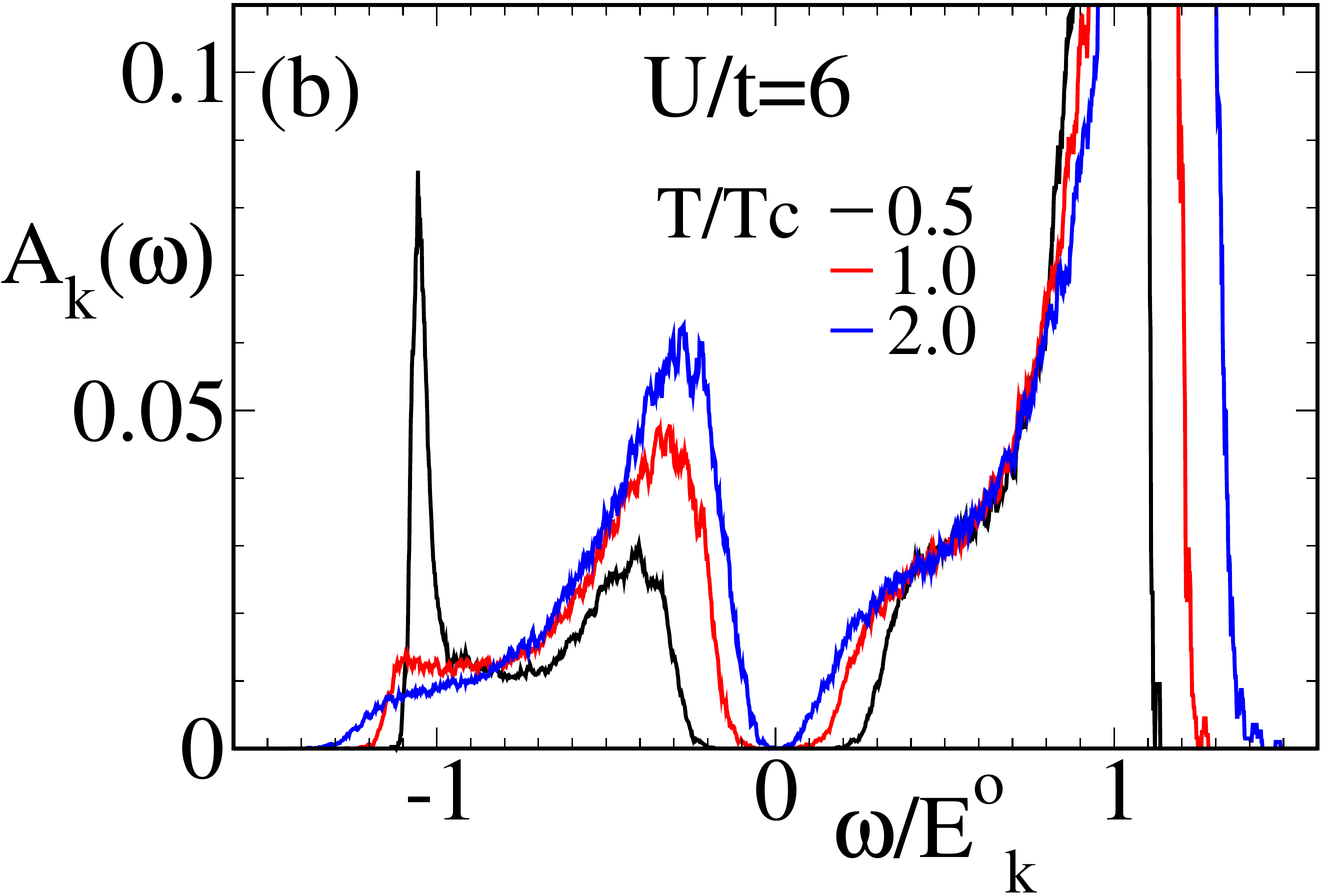}
\includegraphics[width=5.5cm,height=3.9cm,angle=0]{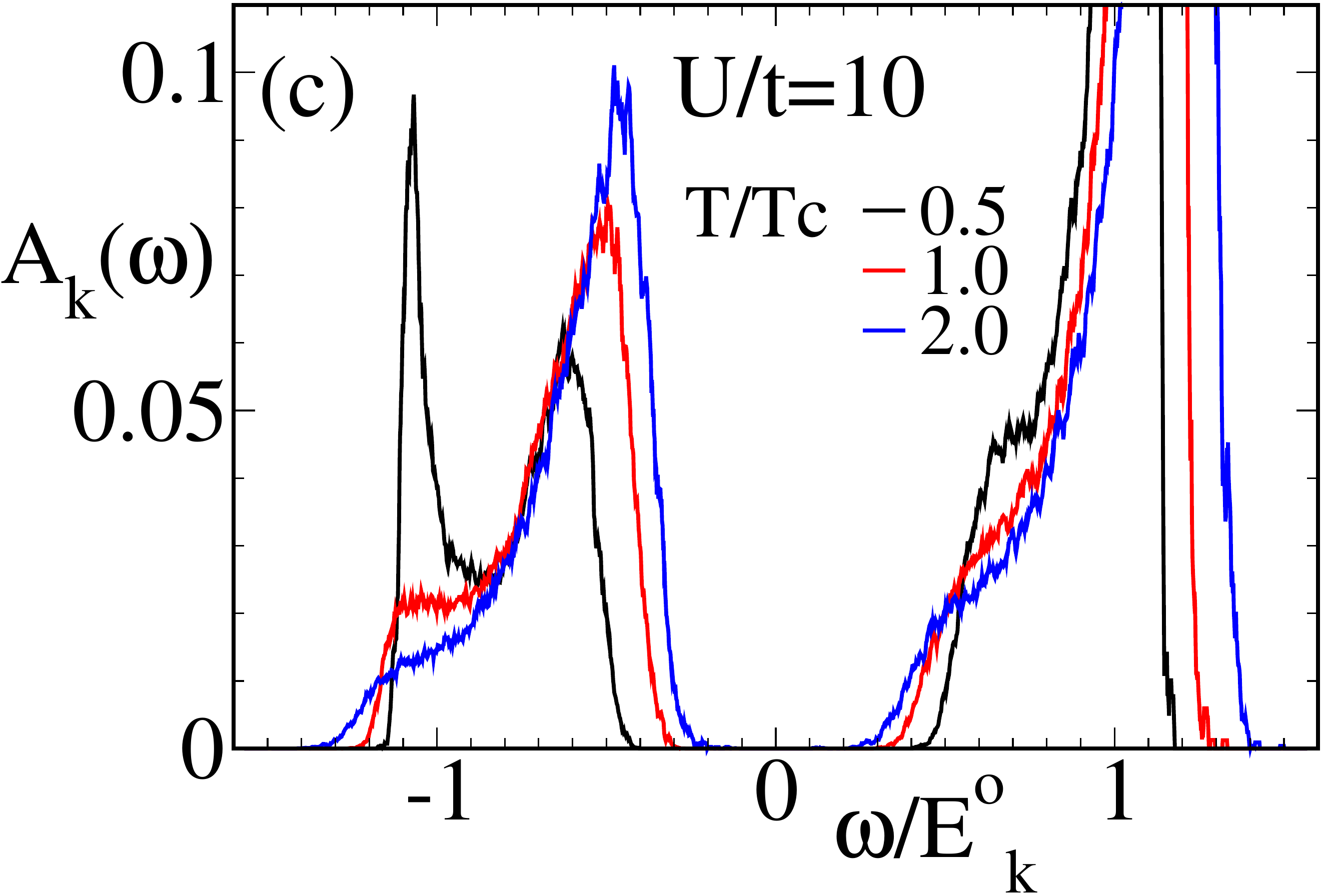}
}
\centerline{
\includegraphics[width=5.5cm,height=3.9cm,angle=0]{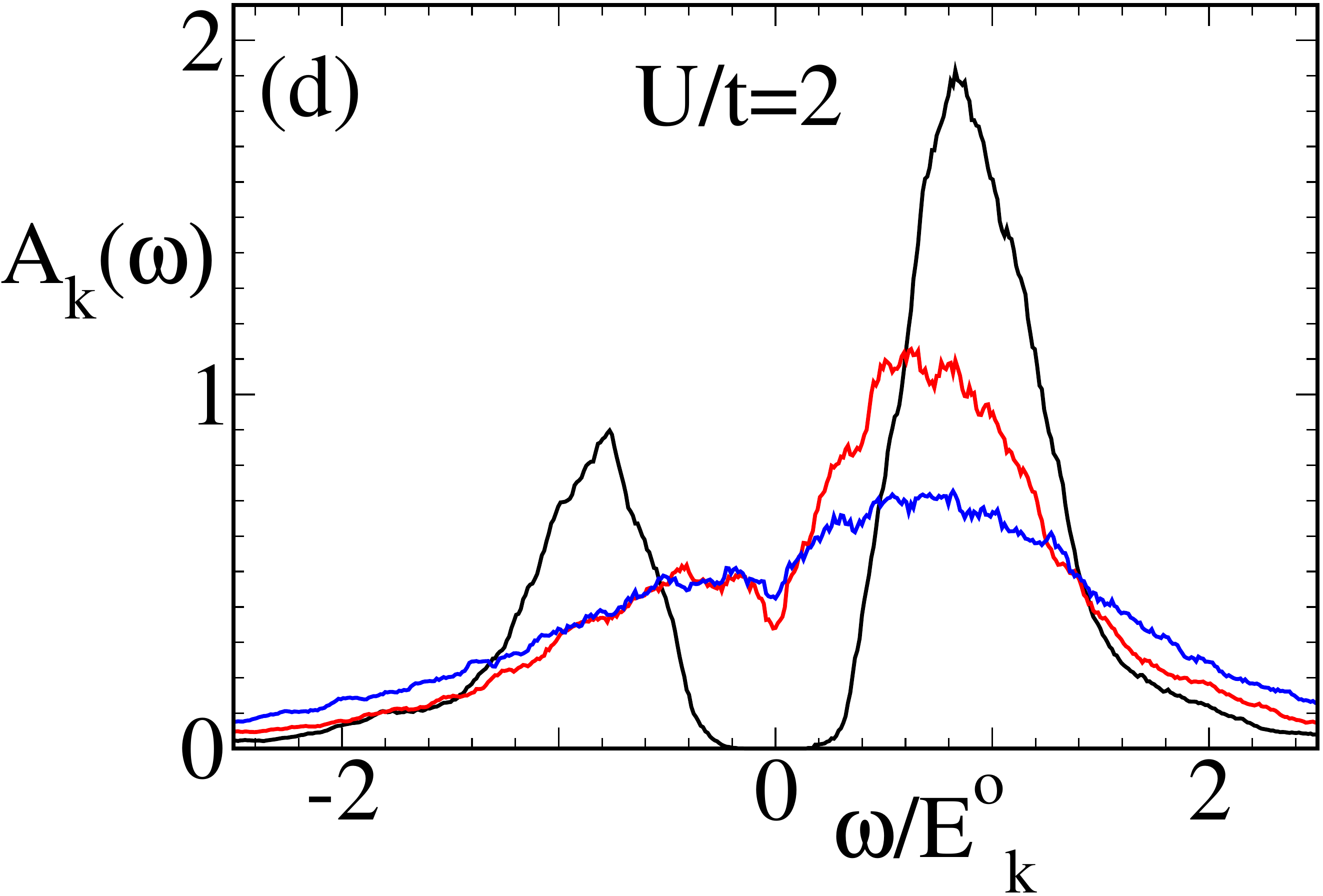}
\includegraphics[width=5.5cm,height=3.9cm,angle=0]{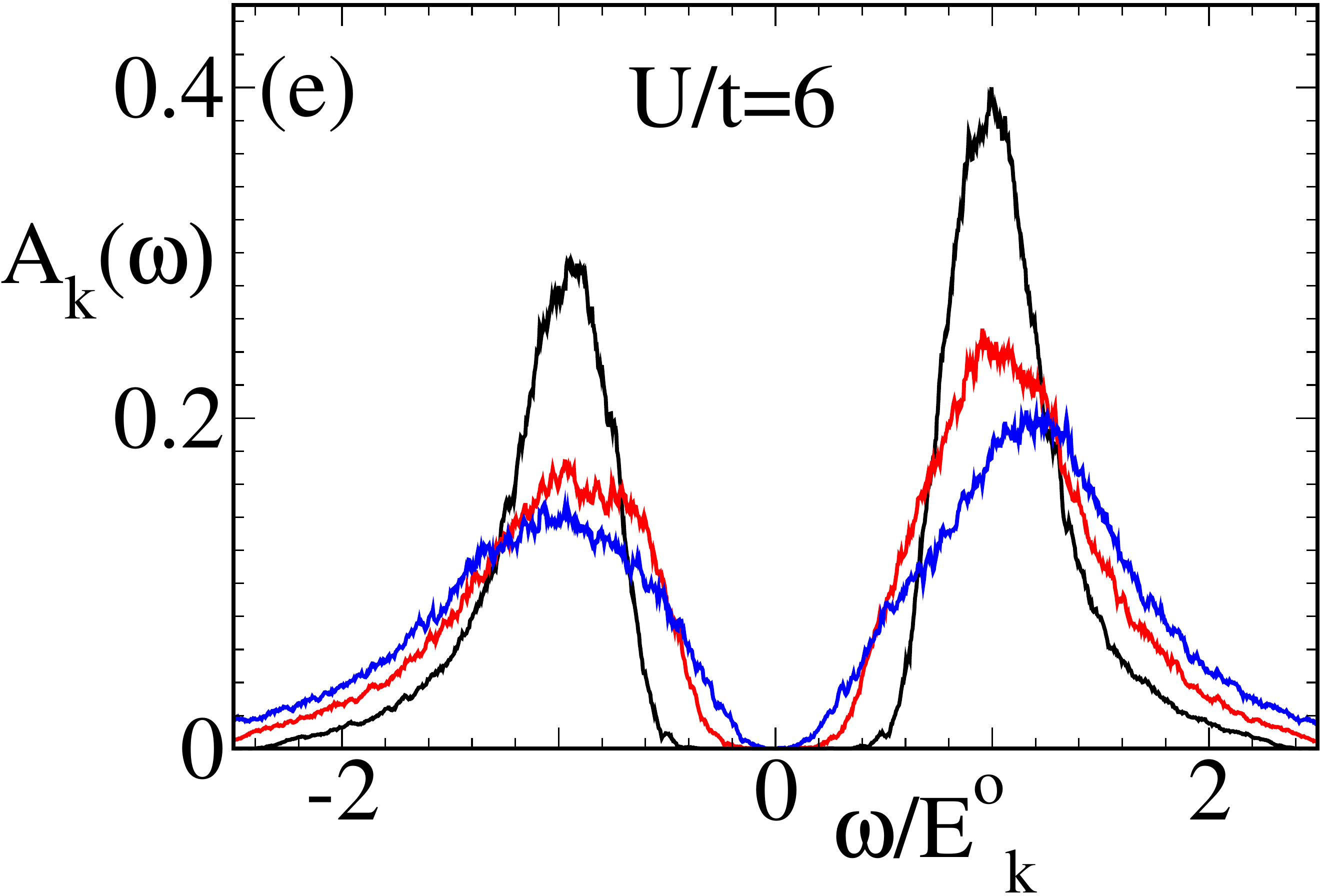}
\includegraphics[width=5.5cm,height=3.9cm,angle=0]{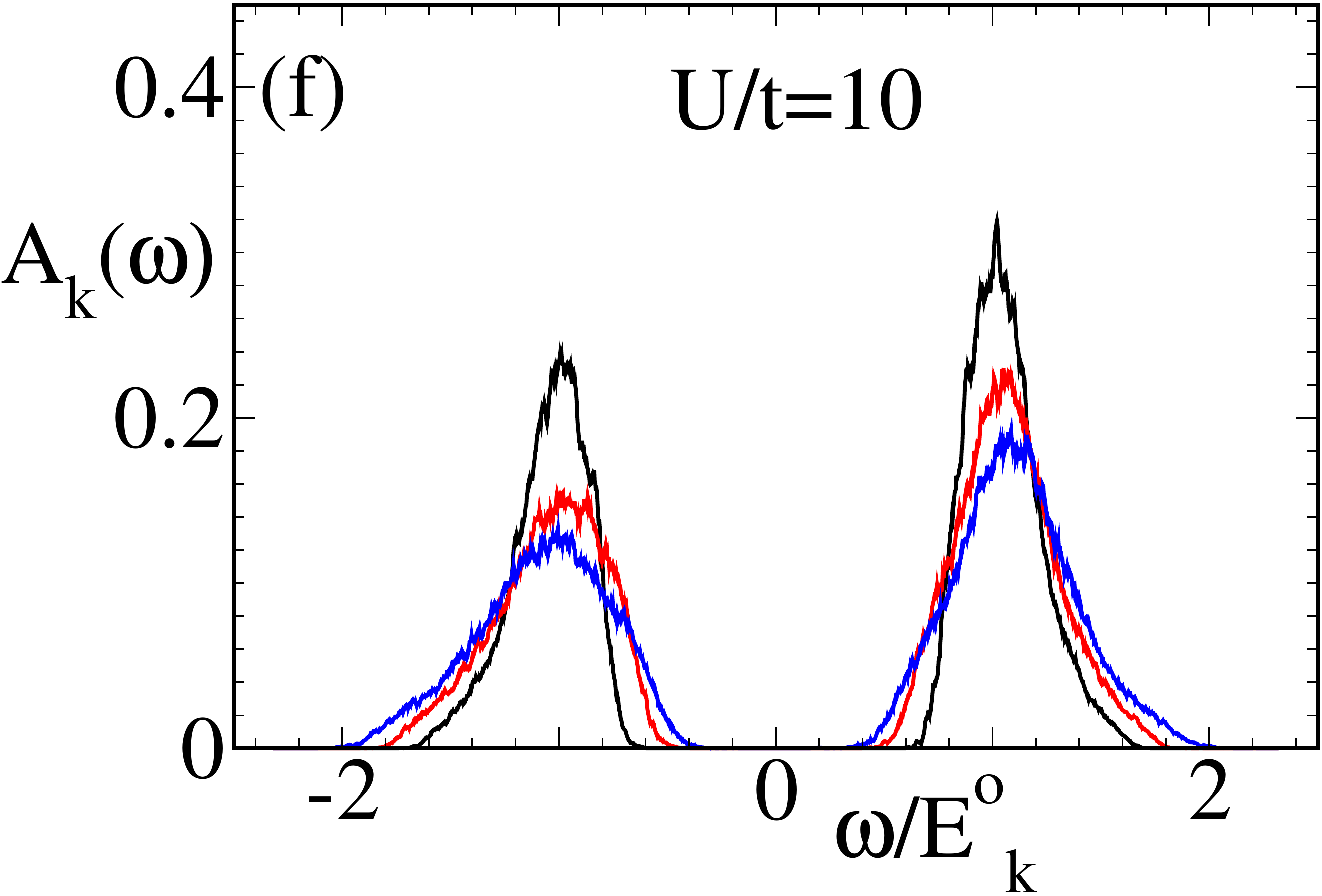}
}
\caption{Colour online: Spectral function 
$A({\bf k}, \omega)$. The panels in the top
row are for ${\bf k} = \{ \pi, \pi\}$. 
(a).~$U/t=2$, (b).~$U/t=6$, (c).~$U/t=10$.
Bottom row, ${\bf k} = \{ \pi/2, \pi/2\}$, and
interaction strengths: (d).~$U/t=2$, (e).~$U/t=6$ and
(f).~$U/t=10$.
For each $U$ we show data at $T= 0.5T_{c}$, $T_{c}$ and $2T_{c}$.
The frequency axis is normalised by the ${\bf k}$ dependent
mean field energy $E^0_{\bf k}$ at $T=0$.
For ${\bf k} = \{ \pi, \pi\}$, which is outside the non interacting
Fermi surface, the basic
structure consists of large peak at positive energies $\omega 
\sim E^0_{\bf k}$, a broad negative energy feature at
$ \omega \gtrsim -E^0_{\bf k}$, and for $T < T_c$ a remnant
of the QP peak at $\omega = -E^0_{\bf k}$. 
Beyond weak coupling the survival of a two peak structure 
even for $T > T_c$ indicates `incoherent pairs'.
For ${\bf k} = \{ \pi/2, \pi/2\}$ the features are similar to
what we observe at ${\bf k} = \{ \pi, \pi\}$, except the 
QP peak is no longer separately visible.
}
\end{figure*}
%------------------------------------------------------------------------

Now we turn to the spectral functions. 
Fig.5 shows intensity plots of the spectral 
function $A({\bf k}, \omega)$ for $U/t=2,~6,~10$
(top to bottom) for $T/T_c=0.1,~1,~2$ (left to right). 
Let us start with the low temperature results, where
mean field theory is a good starting point.
 
Our density $n \sim 0.9$ involves a non interacting Fermi
surface that is almost a square and rotated 
by $45^{\circ}$ with respect to the Brillouin zone.
So, the separation between the two branches, $\pm
\sqrt{(\epsilon_{\bf k} - \mu)^2 + \Delta_0^2}$,
of the mean field dispersion is smallest 
near ($0$,$\pi$) and ($\pi/2$,$\pi/2$).
Within MFT the
spectral function is given by
$A({\bf k}, \omega) = 
u^{2}_{\bf k} \delta(\omega - E_{\bf k}) + 
v^{2}_{\bf k} \delta(\omega + E_{\bf k})$.
In the BCS limit, $v^{2}_{\bf k}$ is either one 
or zero for $k< k_{F}$ or $k > k_{F}$,
with a small region around $k_{F}$ where it 
crosses from one to the other. 

At $U=2t$, $v_{\bf k}^2 \sim 1$ for 
$\epsilon_{\bf k} \lesssim  \mu$,
and $u_{\bf k}^2 \sim 1$ for 
$\epsilon_{\bf k} \gtrsim \mu$, with
significant mixing only near $\mu$.
As a result $A({\bf k}, \omega)$ 
shows either a lower branch or an upper
branch, but not both - except for
$\epsilon_{\bf k} \sim \mu$.

The growing symmetry in the plots,
about the horizontal $\omega=0$ line, 
with increasing $U/t$ arises from the 
changing character of $u_{\bf k}$ and 
$v_{\bf k}$.
For $U/t \gg 1$, $u_{\bf k}^2$ and
$v_{\bf k}^2$ are both $\sim 1/2$ all
over the Brillouin zone, since `pairing'
is no longer limited to the vicinity of
the non interacting Fermi surface (FS).
The cases $U=6t$ and $U=10t$ are already in
this regime although some residual
asymmetry is visible.
The large difference between weak and strong coupling 
in terms of the $T=0$
pairing amplitude decides the finite $T$ state.

At finite $T$, thermal fluctuations broaden the 
delta functions and the detailed lineshapes for 
${\bf k} = \{ \pi, \pi\}$ and $\{ \pi/2, \pi/2\}$ are
shown in Fig.6.  As expected, the gap closes
for $U=2t$, while it does not for $U=6t$ and $U=10t$, 
though there is a noticeable decrease 
in the former.

Quantum Monte Carlo work \cite{qmc-singer} had
suggested the presence of  
non-trivial structure in the spectral 
function near the zone boundary 
($ \pi, \pi $). 
The top row in Fig.6
shows the spectral function at this
${\bf k}$ point for $U/t=2,~6,~10$
at three temperatures, 
$T=0.5T_{c},~T_{c}, ~2T_{c}$.  
The energy is measured in units of the $T=0$ dispersion
$E^0_{\bf k}$. 

We start with the top row: ${\bf k} = \{ \pi, \pi\}$.
At high temperature 
the $U=2t$ case shows only a single broad 
peak at positive energy, whereas
$U/t=6,~10$ show a second peak at a smaller 
negative energy value. 
This two peak structure with a gap around $\omega=0$ 
is an indicator of pairing without
global coherence. The complete absence at 
$U=2t$, and the increase in peak height from 
$U=6t$ to $10t$ bolsters this interpretation. 

We do not find any peak 
near $\omega \sim 0$ for medium to large 
$U$. However from $T_{c}$ downwards, 
another peak becomes visible at negative 
energies, at $\omega \sim - E^0_{\bf k}$.
This peak is 
indicative of the global coherence setting in
below $T_{c}$. As $T$ is decreased, 
this peak slowly gains weight while weight
in the `pairing feature'
becomes smaller, with its maximum 
shifts to larger negative energies 
as the gap becomes larger.

Thus, we find some degree of consistency 
with the QMC work
which mainly deals with temperatures 
larger than $T_{c}$, and also find another peak,
indicative of global coherence, that 
starts to develop below $T_{c}$.

For ${\bf k} = \{ \pi/2, \pi/2\}$ and high
temperature the spectral functions at  $U=6t$ and $U=10t$ 
have a gap at $T > T_c$, as before, while the
$U=2t$ result is gapless. In contrast to ${\bf k} = \{ \pi, \pi\}$
however we cannot disentangle the lower temperature
coherence feature, at $\omega = \pm E^0_{\bf k}$ 
from the overall broad band.

\section{Discussion}

Our model has been set up with the explicit 
constraint that it reproduce the standard mean 
field (or HFBdG) result at $T=0$.
It ignores quantum fluctuations of the pairing field.
The impact of these fluctuations 
have been discussed using 
DMFT by Garg et. al. \cite{dmft-garg} and 
Bauer and Hewson \cite{dmft-bauer}.
They find that the {\it qualitative} results for 
the order parameter, spectral gap,
occupation probability and superfluid 
stiffness are all given correctly by the 
mean field method, though it tends 
to overestimate the spectral gaps and
order parameter values at intermediate coupling.
For most of the $U$ window, however, the mean field
results are reasonable. The results from our method
should get better at finite temperature as thermal 
fluctuations become  more important than
zero point quantum fluctuations. A comparison of our 
$T_c$ with QMC estimate bears this out. 

{\it Accuracy of $T_c$ estimate:}
Fig. 7(a) compares our $T_{c}$ with different methods. 
We find that our 
results compare well
with QMC and sophisticated semi-analytic 
methods 
with a slight underestimate 
at medium to large coupling. 
Fig.8(b) shows the size dependence of the $T_{c}$
estimate with data for $L=8,~16,~24$. We see that while
the `critical temperature' decreases noticeably 
from $L=8$ to $16$, it does not change significantly beyond 
$L=16$.
Thus, the estimate we obtain at $L=24$ should be a fair 
approximant to the bulk $T_c$.

% -------------------------------------------------
\begin{figure}[b]
\centerline{
\includegraphics[width=4.2cm,height=4.7cm,angle=0]{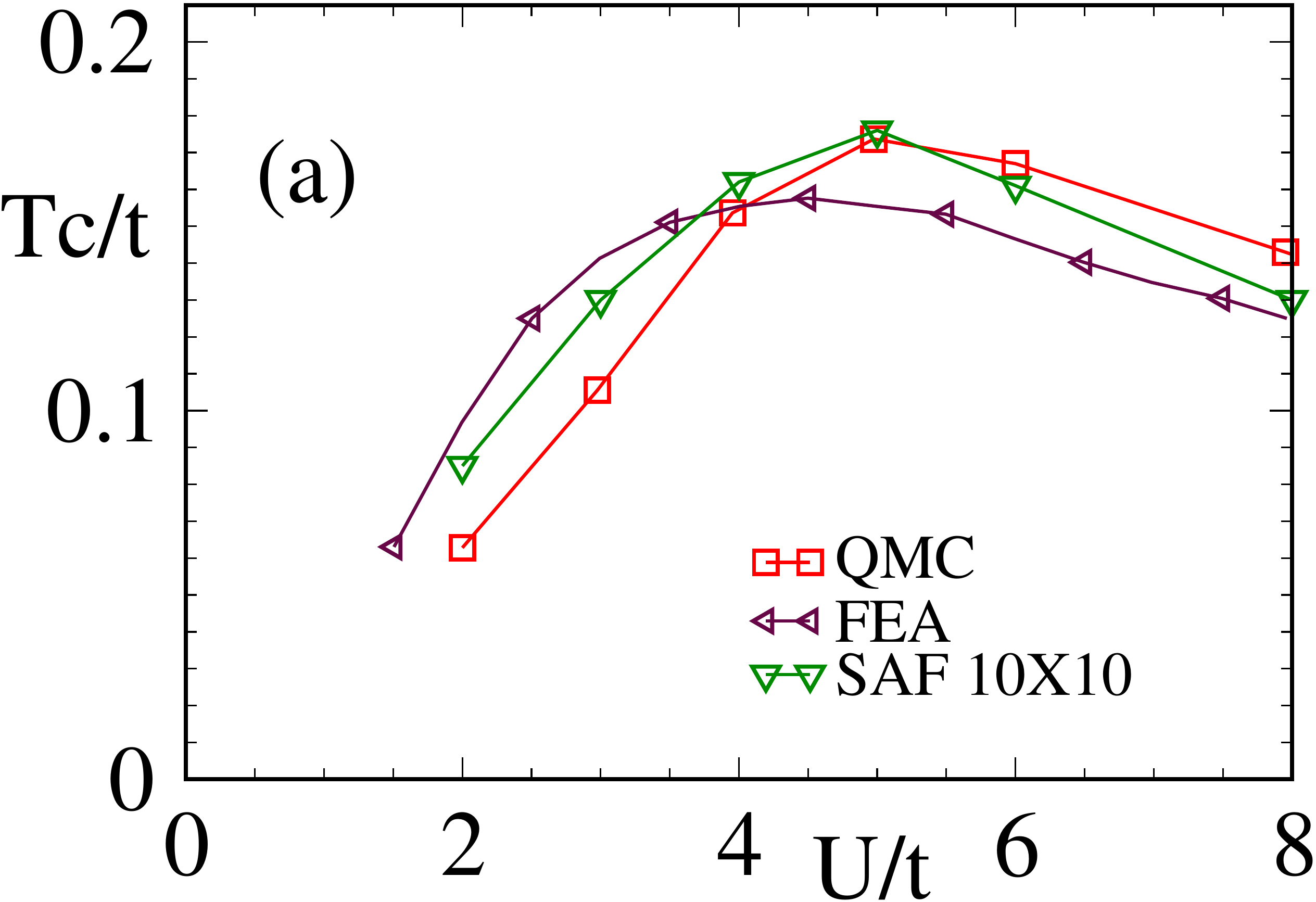}
%\hspace{.1cm}
\includegraphics[width=4.2cm,height=4.7cm,angle=0]{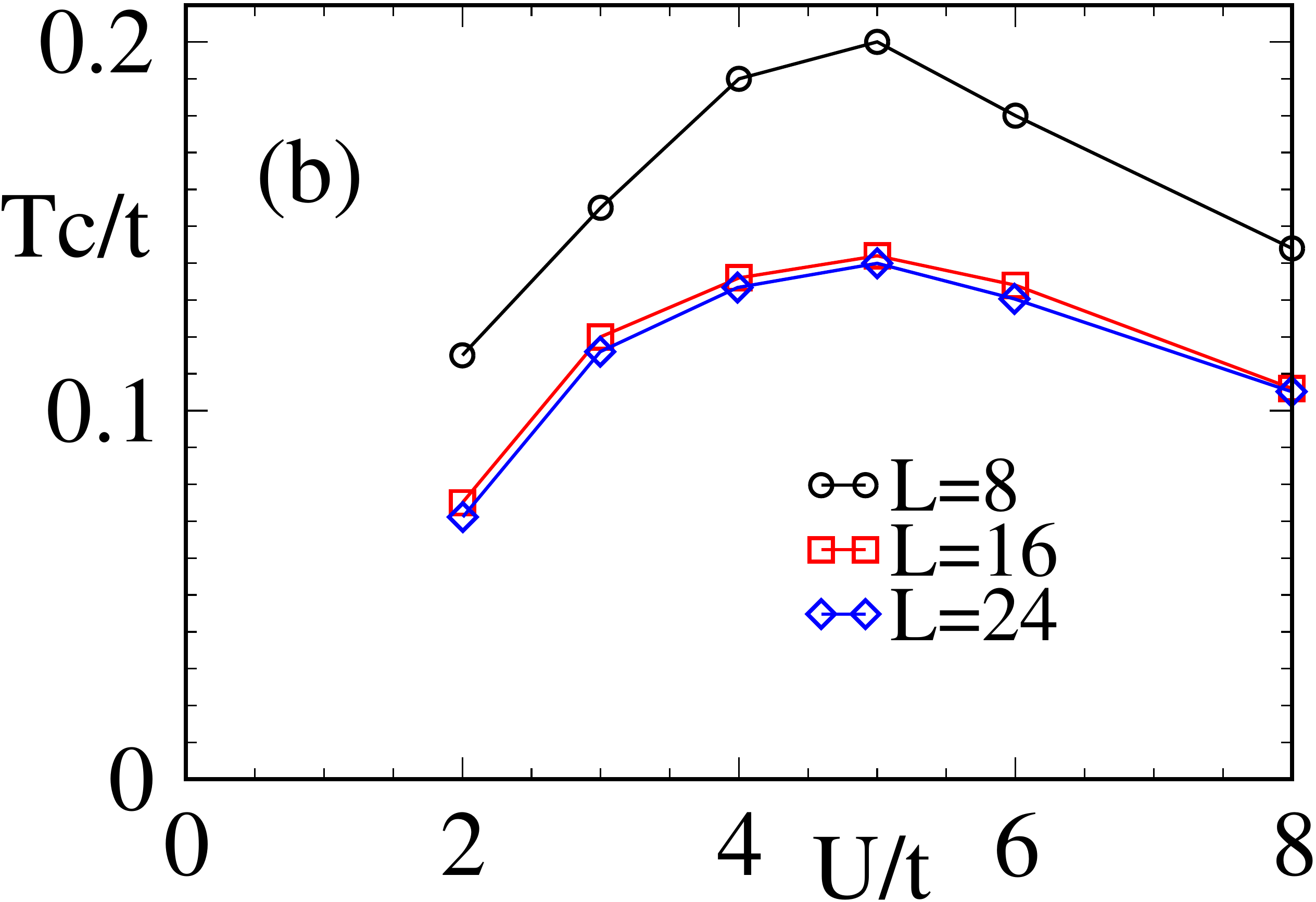}
}
\caption{Colour online: (a) Comparison of
our $T_{c}$ (labelled SAF) with QMC on a 10 $\times$ 10 lattice,
and the semi-analytic method employing the fluctuation exchange approximation
 (FEA) \cite{sa-deisz} . DMFT results \cite{dmft-keller} 
overestimate the $T_{c}$ significantly, and also the location
of peak $T_c$, 
and have not been included in the same plot. (b) Size dependence
of our result, showing that the $T_{c}$ estimate is almost 
size indepenedent after $L=16$.
}
\end{figure}
% -------------------------------------------------

{\it Effective classical functional:}
The BdG framework involves fermions coupled to the fields
$\Delta_i$ and $\phi_i$.  For simplicity let us focus on
the $\Delta_i$ since the $\phi_i$ do not play a crucial
role in a translation invariant system.

The physics of fermions in an arbitrary $\Delta_i$ 
background is not obvious. It is therefore helpful
to have an explicit classical functional involving
only the $\Delta_i$ since the minimum and possible
fluctuations in $\Delta_i$ are easier  to estimate. 

If the $\Delta_i$ are small compared to the kinetic
energy, as would happen when $U/t \ll 1$, the 
functional, $H^0_{eff}$, can be obtained via a
standard cumulant expansion: 
$$
H^0_{eff}\{\Delta_i\} = \sum_{ij} a_{ij} \Delta_i \Delta^*_j
+ \sum_{ijkl} b_{ijkl} \Delta_i \Delta^*_j \Delta_k \Delta^*_l
+ {\cal O}(\Delta^6)
$$
The superscript in $H_{eff}$ is to indicate $U/t \ll 1$ 
character.  
$a_{ij} = -\chi^0_{ij}/2 + (1/U)\delta_{ij}$, $\chi^0_{ij}$
being the non-local pairing susceptibility of the free Fermi
system, and $b_{ijkl}$ can be computed from a convolution of
four free Fermi Greens functions.

% -------------------------------------------------
\begin{figure*}[t]
\centerline{
\includegraphics[width=4.5cm,height=4.0cm,angle=0]{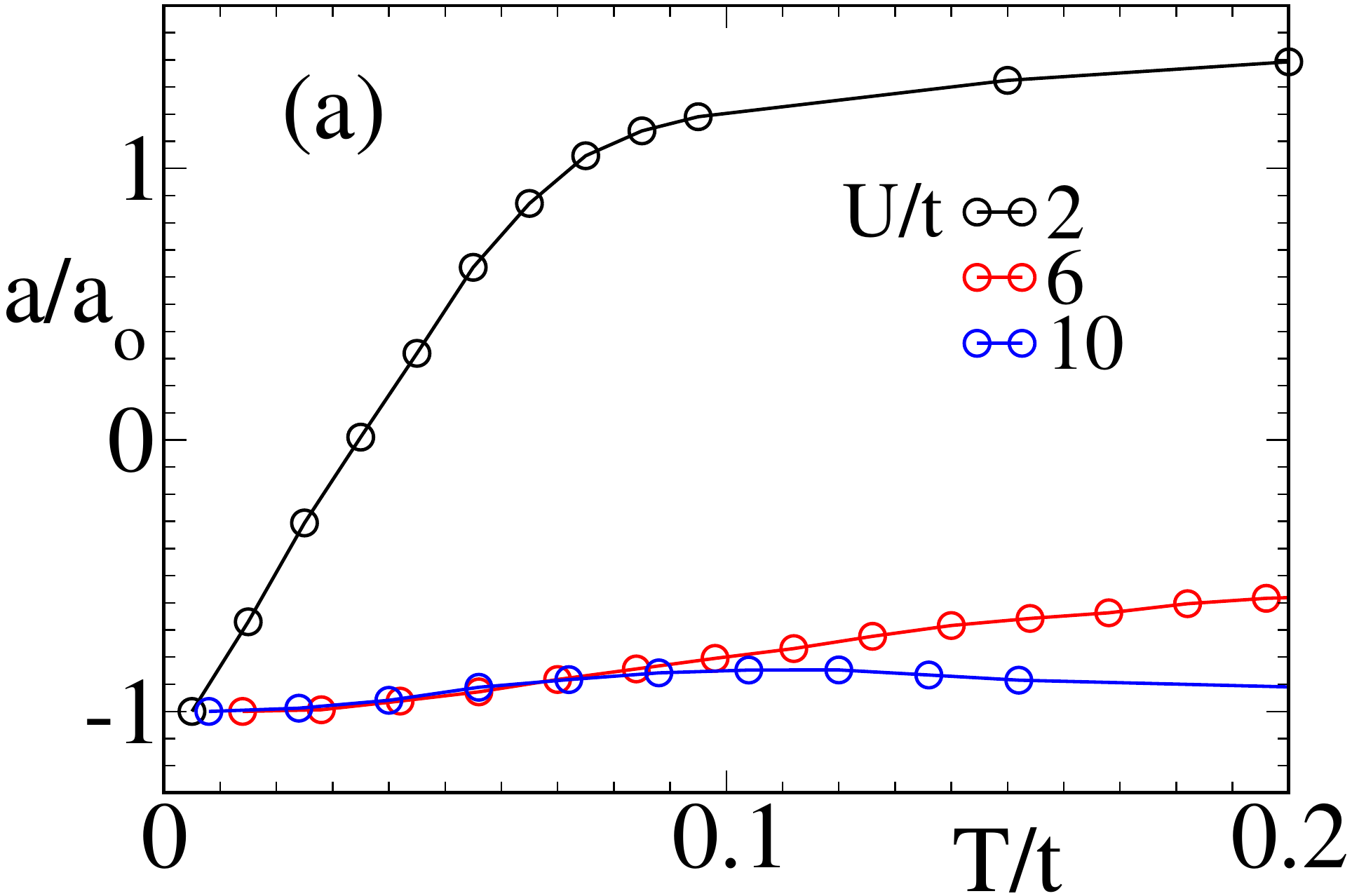}
\hspace{.1cm}
\includegraphics[width=4.5cm,height=4.0cm,angle=0]{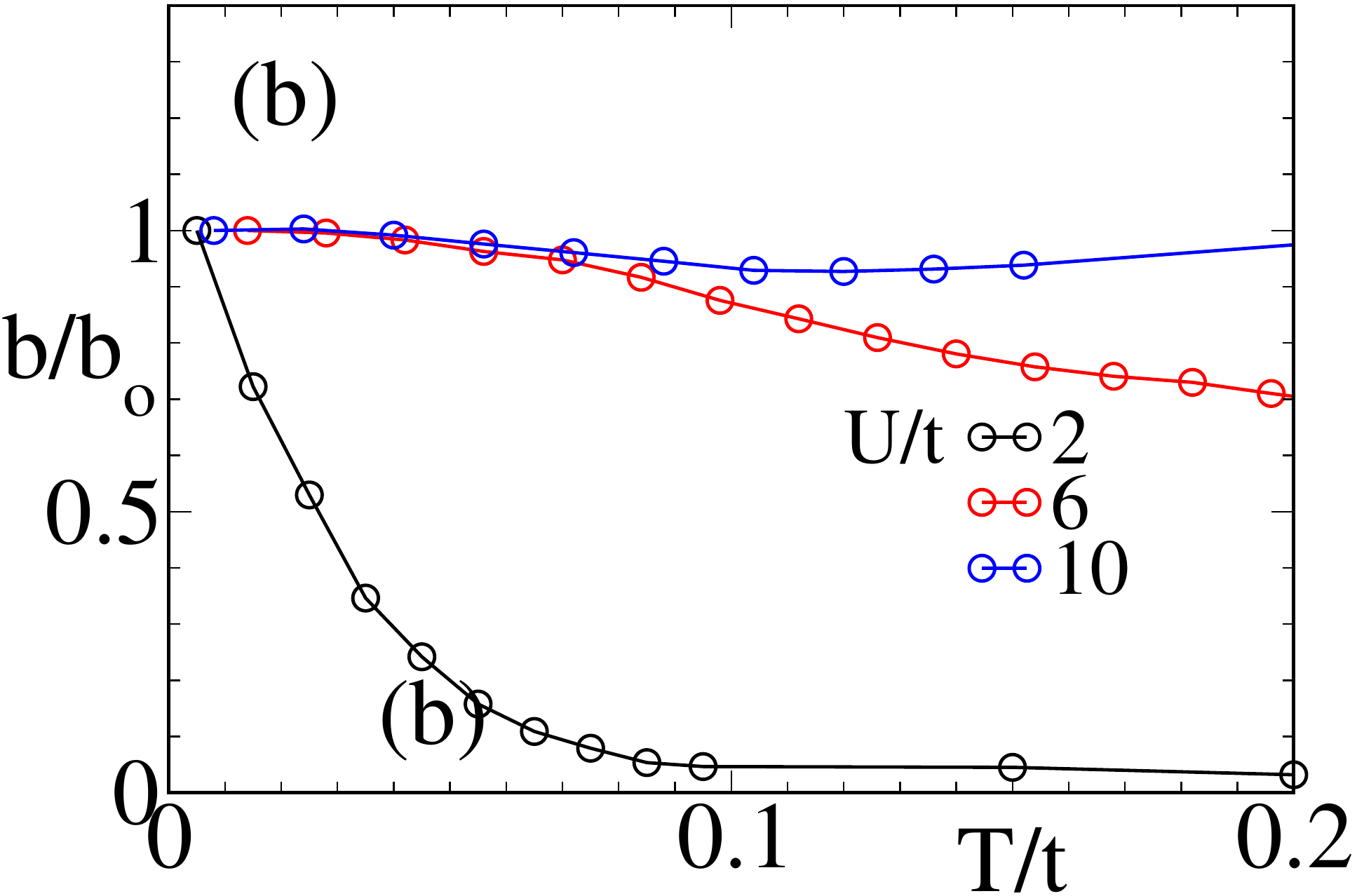}
\hspace{.1cm}
\includegraphics[width=4.5cm,height=4.0cm,angle=0]{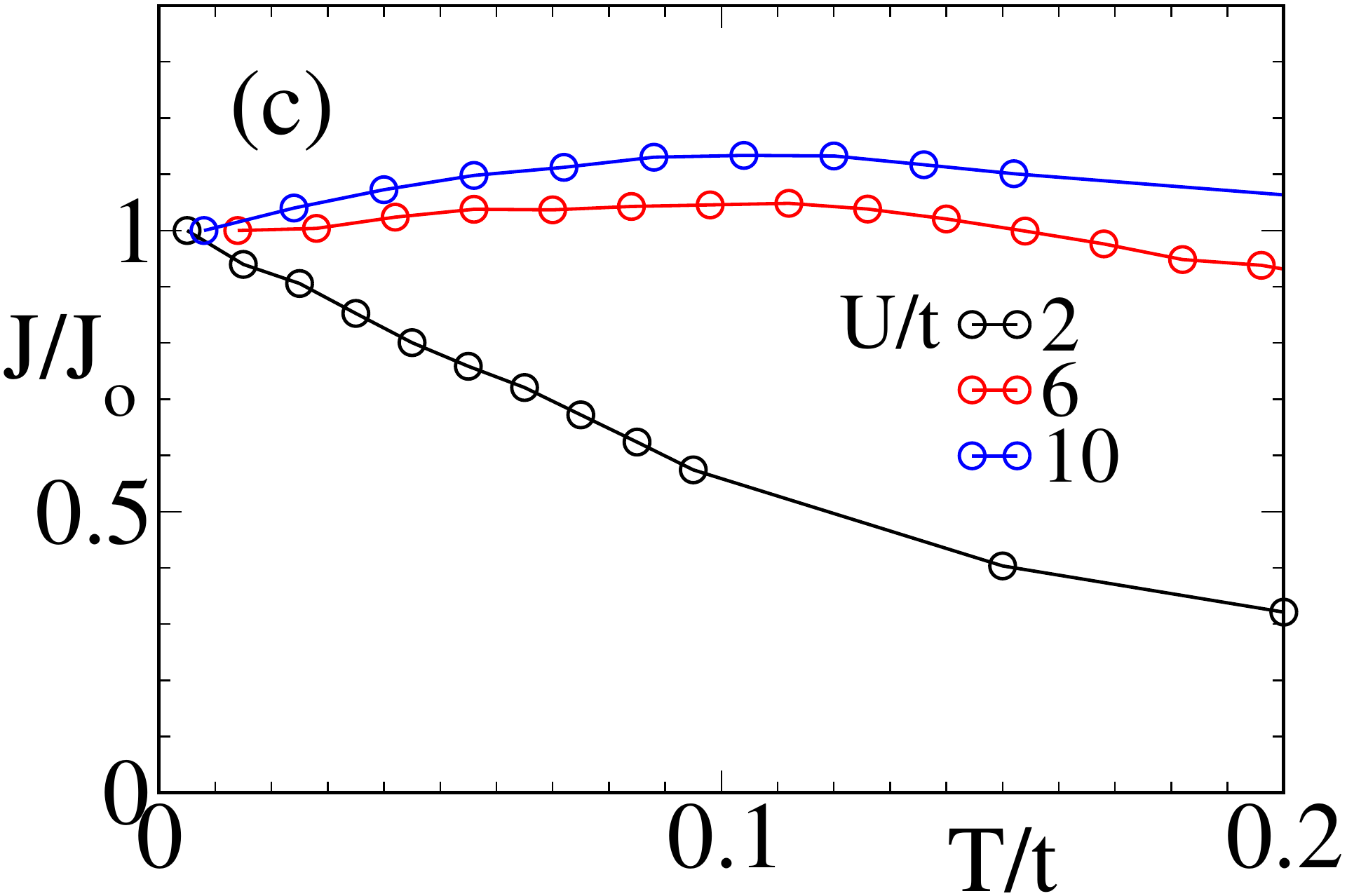}
}
\caption{Colour online: Parameters defining the phenomenological
model. (a)~The parameter $a(T,U)$, (b).~the parameter $b(T,u)$,
and (c).~the stiffness $J(T,U)$ of the effective XY model
for the phase degrees of freedom.
$a$ and $b$ are normalised to their $T=0$ values. Notice the
essential flatness of $a(T)$ and $b(T)$ at $U=10t$, the weak
$T$ dependence at $U=6t$, and the dramatic variation with
$T$ at $U=2t$. $J$ similarly is only weakly $T$ dependent
for $U \gtrsim 6t$ and varies strongly with $T$ at 
weak coupling.
}
\end{figure*}
% -------------------------------------------------

If we had $U/t \gg 1$
then $H_{eff}$ would have to be expanded to higher order
in $\Delta_i$. In that situation it actually helps to extract the
functional by expanding in powers of $t/\Delta$, leading to
the strong coupling limit:
$$
H^{\infty}_{eff} \{ \Delta_i \} \approx -t^2/{\vert \Delta \vert}  
\sum_{\langle ij \rangle} 
cos(\theta_i - \theta_j) + \sum_i H_{loc}(\vert \Delta_i \vert)
$$ 
$H_{loc}(\vert \Delta_i \vert)$ can be obtained from the
atomic problem. The leading intersite term is calculated
perturbatively and connects only nearest neighbour sites.

While these BCS and BEC limits are easy, obtaining an usable
functional at arbitrary $U/t$ does not seem possible. We have
therefore tried a parametrisation of the (local) amplitude
fluctuation spectrum and the phase correlations in terms 
of the following phenomenological model. It is valid at all $U/t$
and over the temperature window of interest.
$$ 
H^{phen}_{eff} = -\sum_{\langle ij \rangle} J cos(\theta_i-\theta_j)
+ \sum_{i} \{ a |\Delta_{i}|^{2} + b |\Delta_{i}|^{4} \}
$$
The first term defines an effective XY model involving only
the phases, but, as we will see, the $J$ needs to be
temperature dependent to incorporate the effect of amplitude
fluctuations.
The amplitude part of $H_{eff}$ is purely local, and to that
extent misses out on spatial correlation between amplitude
fluctuations. 

The parameters $a$ and $b$  are
extracted from a fit to the $P(\vert \Delta \vert)$ that we 
obtain from the full MC, see Fig.3. 
With the moments of $\vert \Delta_i \vert$ fixed by $a$ and
$b$, the $J(U,T)$ is 
obtained by imposing the following equality:
$$
\langle \sum_{ij} \vert \Delta_i \vert \vert
\Delta_j \vert cos(\theta_i - \theta_j) \rangle_{MC}
= 
\langle 
\sum_{ij} 
\vert \Delta^2 \vert  
cos(\theta_i - \theta_j) \rangle_{phen}
$$
The left hand side is the MC based order parameter,
Fig.1(a). The right hand side computes the same
quantity within the phenomenological model (in which
the $\vert \Delta \vert$
and $\theta$ averages factorise) by using a Monte Carlo
estimate of $\langle cos(\theta_i - \theta_j) \rangle$
in the XY model.

Fig.8.(a)-(b) 
shows the $T$ dependence of $a$ and $b$ for $U=2$,$U=6$ 
and $U=10$. Because of the large difference in 
scales between the weak and strong coupling 
we have normalized the parameters by their $T=0$ values. 

The $U=2t$ parameters show large change with $T$. The normalized
$a$ quickly increases and becomes positive, while $b$ rapidly 
decreases from its $T=0$.
Both parameters tend to saturate for $T \sim T_{c} \sim 0.07t$.
The 6th order term in the expansion would be necessary to describe
the $U=2t$ case accurately.

With increasing $U$, the thermal change of the parameters 
slows down, and even at $U=6t$ the parameters
show a much weaker dependence on $T$. By $U=10t$, they are 
essentially constant at their $T=0$ values, indicating that 
only phase fluctuations are relevant in this regime.

To gain more perspective, we consider an expansion of the distribution
about its mean value, 
$P(\Delta) =  K_2 (\Delta-\Delta_0)^2 + 
K_3 (\Delta-\Delta_0)^3 + K_4 (\Delta-\Delta_0)^4 ..$,
where the first term represents the gaussian stiffness of the 
distribution and the other 
terms represent non gaussian contributions. 
At strong coupling, the physics is 
driven completely by the phase fluctuation term, 
and the magnitude of the amplitudes
is almost fixed. Thus, the stiffness coefficient is very large. 
As the coupling decreases, amplitude fluctuations increase, 
hence signalling a decrease in the stiffness.  
Apart from the increase in amplitude fluctuations, 
the mean value of $\Delta$ also
shows a remarkable increase with $T$ at weak 
coupling, signalling the importance
of the non gaussian terms in the expansion.
We next turn to 
examine the phase stiffness  which is the crucial
coupling at large $U$.

Fig.8.(c) shows the $T$ dependence of $J$ for the three couplings, 
again normalized by their $T=0$ values, $J_{0}(U)$.
The $U=2t$ case shows a 
pronounced decrease with $T$, while the other two are effectively
constant.
An XY description with a $T$ independent coupling is reasonable
for $U=6t$ and $10t$ but inadequate at $U=2t$.

{\it Role of the `density' field:} 
An important addition in our model is the field  
$\phi$, coupling to the density
operator. It serves a twofold purpose: first, it is 
indispensable in a disordered system
since it provides a site dependent background
 field that renormalises the total disorder,
and is crucial to get the correct scales; and second, 
it incorporates fluctuations
in the charge sector, which play an important role 
for $n \sim 1$. 

At $n=1$ the negative $U$ Hubbard 
model can be mapped to its positive $U$
counterpart, with the 
components of the
magnetization field, ${\bf m}_i$ of the positive 
$U$ model
corresponding to the $\Delta_i$ and $\phi_i$.
The symmetry 
the model is increased from $O(2)$ to 
$O(3)$, and hence, there can be no 
superconducting order at finite temperature
in 2D.
At  $T=0$, the superconducting state is degenerate
with the charge density wave state. This degeneracy 
is built into the structure of our model,
and simulations at $n \sim 1$ do actually show both 
superconductivity and charge density wave order
at low $T$. 
The two field decomposition captures 
the correct ground state and relevant fluctuations
in the model.

{\it Handling inhomogeneity:}
As remarked earlier, this method is particularly
well suited to dealing with inhomogeneous systems,
including disordered systems and systems in a trap.
We have extensively studied both of these,
the former in the context of the disorder induced
superconductor-insulator transition \cite{trans,spat}, and
the latter in the context of superconductors 
in a harmonic trap \cite{catom}. In such inhomogeneous
systems, the Hartree feedback plays a crucial role
in modifying the effective potential that the 
electronic system sees. In the former, this is crucial in 
determining the correct critical disorder and 
moderate disorder charge transport properties 
\cite{trans}, and plays a major role in the spatial
fragmentation of the system \cite{spat}. In a harmonic trap,
the resultant inhomogeneous density profile
can drastically alter the spectral properties
of the system \cite{catom}, compared to a flat one.
Similarly for FFLO phases in imbalanced Fermi
systems the real space treatment on large
lattices allow access to a wealth of non trivial
modulated phases.

{\it Quantum fluctuations:}
The major approximation in our model 
is the neglect of temporal fluctuations in the
auxiliary fields. The primary effect 
is the absence of low energy `bosonic'
modes (due to preformed pairs) at strong coupling. 
This does not affect the thermodynamics and 
single particle spectrum significantly.
Two particle 
correlations like conductivity also give accurate
results as long as we are at moderate $U$.
However, as we increase $U/t$
the system develops a pseudo-gap (or a gap)
above $T_{c}$, and we get a
resistivity with steadily increasing 
insulating character. In the complete treatment the
bosonic modes allow a parallel channel of conduction.
A purely static approximation misses this contribution 
at large $U/t$, as does DMFT. 

\section{Conclusions}

We have presented results on BCS-BEC crossover in an
attractive Fermi system in the context of 
the two dimensional
Hubbard model. We use an auxiliary field 
decomposition, treat these
fields as classical, and solve the resulting problem through
a real space Monte Carlo technique.
The inclusion of all spatial thermal fluctuations allows
us to capture the correct $T_c$ all the way from the BCS
to the BEC end. It allows conceptual clarity about the 
amplitude and phase fluctuation dominated asymptotes and
the crucial intermediate coupling window where both these
fluctuations are relevant. We provide a detailed 
characterisation of the auxiliary field behaviour that
dictates fermion physics and access results on the
density of states and angle resolved spectral features
without any need for analytic continuation.
We lay the groundwork for
the study of disordered superconductors, trapping effects in
superfluids, and spontaneous inhomogeneity in imbalanced
systems.

{\it Acknowledgments:}
We acknowledge use of the High Performance Computing Cluster 
at HRI.  PM acknowledges support from a DAE-SRC 
Outstanding Research Investigator Award.

\end{document}